\documentclass[traditabstract]{aa}
\usepackage{graphicx}
\usepackage{txfonts}
\usepackage{natbib}
\bibpunct{(}{)}{;}{a}{}{,}

\newcommand{\Apx}{\,\AA\,px$^{-1}$}                        
\newcommand{\apx}{$^{\prime\prime}$\,px$^{-1}$}            
\newcommand{\Porb}{\ensuremath{P_{\rm orb}}}               
\newcommand{\mc}[1]{\multicolumn{2}{c}{#1}}
\newcommand{\reff}[1]{{#1}}                                

\title{Orbital periods of cataclysmic variables identified by the SDSS.}

\subtitle{IX. NTT photometry of eight eclipsing and three magnetic systems}

\author{John Southworth \inst{1} \and C.\ Tappert \inst{2} \and
        B.\ T.\ G\"ansicke \inst{3} \and C.\ M.\ Copperwheat \inst{4}}

\institute{Astrophysics Group, Keele University, Staffordshire, ST5 5BG, UK \ \ \ \ \ \email{astro.js@keele.ac.uk}
           \and Instituto de Astronom\'{\i}a y Astrof\'{\i}sica, Pontificia Universidad Cat\'olica, Av.\ Vicu\~na Mackenna 4860, 782-0436 Macul, Santiago, Chile
           \and Department of Physics, University of Warwick, Coventry, CV4 7AL, UK
           \and Astrophysics Research Institute, Liverpool John Moores University, Liverpool, L3 5RF, UK
          }

\date{Received 2014/09/26; accepted 2014/11/10}       

\abstract{We report the discovery of eclipses and the first orbital period measurements for four cataclysmic variables, plus the first orbital period measurements for one known eclipsing and two magnetic systems.
SDSS J093537.46$+$161950.8 exhibits 1-mag deep eclipses with a period of 92.245\,min.
SDSS J105754.25$+$275947.5 has short and deep eclipses and an orbital period of 90.44\,min. Its light curve has no trace of a bright spot and its spectrum is dominated by the white dwarf component, suggesting a low mass accretion rate and a very low-mass and cool secondary star.
CSS J132536$+$210037 shows 1-mag deep eclipses each separated by 89.821\,min.
SDSS J075653.11$+$085831.8 shows 2-mag deep eclipses on a period of 197.154\,min.
CSS J112634$-$100210 is an eclipsing dwarf nova identified in the {\it Catalina Real Time Transit Survey}, for which we measure a period of 111.523\,min.
SDSS J092122.84$+$203857.1 is a magnetic system with an orbital period of 84.240\,min; its light curve is a textbook example of cyclotron beaming.
A period of 158.72\,min is found for the faint magnetic system SDSS J132411.57$+$032050.4, whose orbital light variations are reminiscent of AM\,Her.
Improved orbital period measurements are also given for three known SDSS cataclysmic variables.
We investigate the orbital period distribution and fraction of eclipsing systems within the SDSS sample and for all cataclysmic variables with a known orbital period, with the finding that the fraction of known CVs which are eclipsing is not strongly dependent on the orbital period.}


\begin{document} \maketitle 

\section{Introduction}                                                                       \label{sec:intro}

Cataclysmic variables (CVs) are interacting binary stars composed of a white dwarf orbited by a low-mass secondary star which fills its Roche lobe. In most CVs the secondary star is hydrogen-rich and loses material to the white dwarf via an accretion disc. Comprehensive reviews on the subject of CVs have been given by \citet{Warner95book} and \citet{Hellier01book}. Their evolution is dominated by the loss of orbital angular momentum, which results in CVs evolving from longer orbital periods down to a period minimum, caused by the changes in the structure of the mass donor, before turing back to longer periods. This evolutionary process leaves a strong imprint on the orbital period distribution of the known CV population, which is discussed in detail by \citet{Gansicke+09mn}.

The optical light of most CVs is dominated by continuum emission and broad emission lines arising from the accretion disc. This swamps the spectral signatures of the white dwarf and secondary star, making it very difficult to characterise the physical properties of these components. Particularly for short-period CVs, the mass donor is often only detectable if it occults the white dwarf and accretion disc. Eclipsing CVs are therefore a valuable resource, as analysis of the eclipse shapes is one of the few ways of revealing the masses and radii of all three components of the system \citep{Wood+89apj,Horne++91apj,Littlefair+06sci,Littlefair+08mn}.

A fraction of CVs harbour magnetic white dwarfs, and these objects have quite different evolutionary processes to the non-magnetic CVs \citep{WebbinkWick02mn,Norton++04apj}. If the  white dwarf's magnetic field is sufficiently strong it disrupts the accretion disc, and accretion occurs along the field lines to the magnetic poles on the white dwarf surface. Interaction between the magnetic fields of the white dwarf and the mass donor also suppresses or reduces the efficiency of magnetic braking \citep{Araujo+05apj}. The interplay of plasmas with strong magnetic fields makes these objects natural laboratories for physics in extreme environments.

\begin{figure} \includegraphics[width=0.48\textwidth,angle=0]{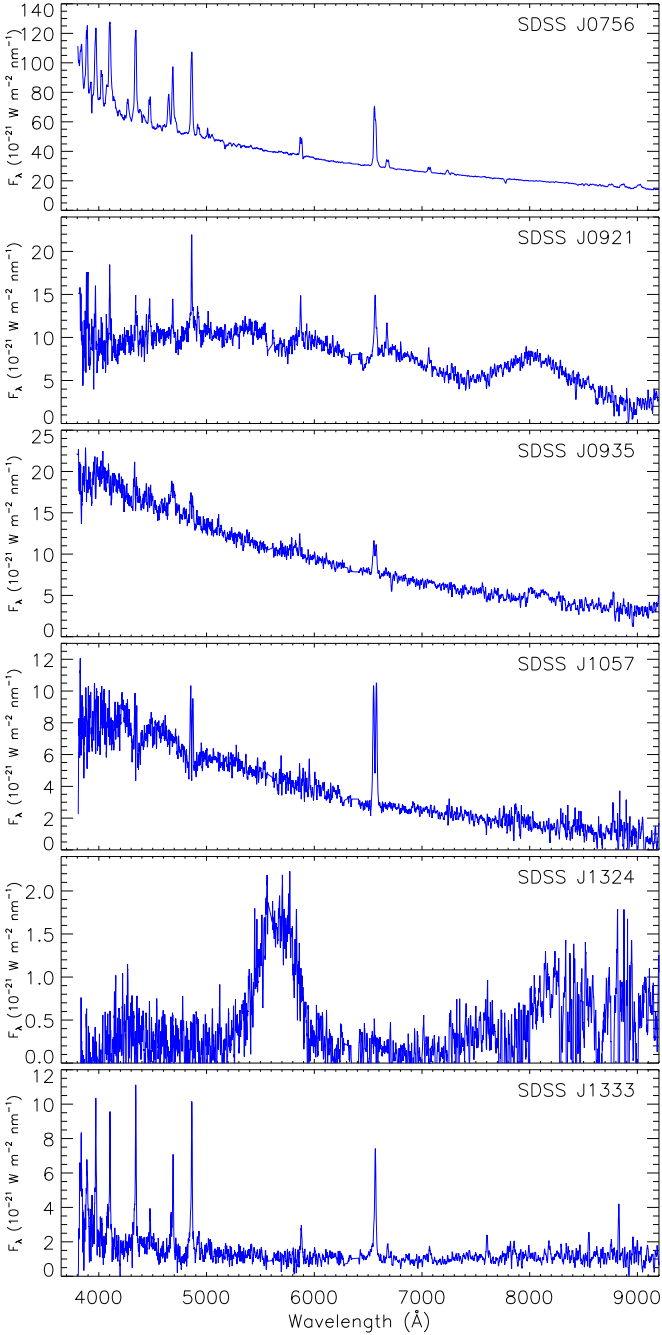} \\
\caption{\label{fig:sdssspec} SDSS spectra of six of the CVs for which we
provide the first precise orbital period measurement (CSS\,J1126 and CSS\,J1325
have not been observed spectroscopically by the SDSS). The flux levels have been
smoothed with 10-pixel Savitsky-Golay filters for display purposes. The units of
the abscissae are $10^{-21}$\,W\,m$^{-2}$\,nm$^{-1}$, which corresponds to
$10^{-17}$\,erg\,s$^{-1}$\,cm$^{-2}$\,\AA$^{-1}$.} \end{figure}

We are engaged in characterising the population of CVs spectroscopically identified by the Sloan Digital Sky Survey (SDSS\footnote{\tt http://www.sdss.org/}; see \citealt{Szkody+11aj} and references therein). Our primary aim is to measure the orbital periods of these objects; further information and previous results can be found in \citet{Gansicke+06mn}, \citet{Dillon+08mn} and \citet{Me+06mn,Me+07mn}. Detecting eclipses is one of the most reliable and straightforward ways of measuring the orbital period of a CV. Similarly, magnetic CVs tend to be highly variable in brightness due to processes such as cyclotron emission. In this work we present photometry of eleven objects and measure precise orbital periods for ten of these systems. Basic information for all the objects observed is compiled in Table\,\ref{tab:iddata}, which contains the abbreviated names which we will use throughout this paper. The SDSS spectra for \reff{six of these} objects are reproduced in Fig.\,\ref{fig:sdssspec} for reference.

\begin{table*}
\caption{\label{tab:iddata} Full and abbreviated names, references and SDSS apparent $ugriz$ magnitudes of the targets.}
\centering
\begin{tabular}{lllccccccc} \hline
SDSS name                  & Short name  & Reference           & $u$ & $g$ & $r$ & $i$ & $z$ & $g_{\rm spec}$  \\
\hline
\object{SDSS J075059.97+141150.1} & SDSS\,J0750 & \citet{Szkody+07aj} & 19.20 & 19.09 & 18.98 & 18.79 & 18.58 & 18.83 \\
\object{SDSS J075653.11+085831.8} & SDSS\,J0756 & \citet{Szkody+11aj} & 16.70 & 16.25 & 16.29 & 16.34 & 16.38 & 17.27 \\ 
\object{SDSS J092122.84+203857.1} & SDSS\,J0921 & \citet{Szkody+09aj} & 20.68 & 19.85 & 19.16 & 19.17 & 19.64 & 19.20 \\
\object{SDSS J092444.48+080150.9} & SDSS\,J0924 & \citet{Szkody+05aj} & 19.49 & 19.25 & 19.26 & 18.70 & 17.88 & 19.36 \\
\object{SDSS J093537.46+161950.8} & SDSS\,J0935 & \citet{Szkody+09aj} & 19.51 & 19.08 & 18.99 & 19.01 & 18.99 & 18.77 \\ 
\object{SDSS J100658.40+233724.4} & SDSS\,J1006 & \citet{Me+09aa}     & 18.46 & 18.31 & 17.93 & 17.51 & 17.14 & 18.63 \\
\object{SDSS J105754.25+275947.5} & SDSS\,J1057 & \citet{Szkody+09aj} & 19.87 & 19.61 & 19.61 & 19.81 & 19.65 & 19.64 \\
\object{CSS J112634-100210}       & CSS\,J1126  & \citet{Drake+09apj} & 18.81 & 18.81 & 18.61 & 18.38 & 18.12 &       \\ 
\object{SDSS J132411.57+032050.4} & SDSS\,J1324 & \citet{Szkody+04aj} & 23.74 & 22.08 & 20.46 & 20.19 & 19.42 & 23.29 \\
\object{CSS J132536+210037}       & CSS\,J1325  & \citet{Wils+10mn}   & 21.94 & 23.10 & 20.38 & 20.63 & 20.55 &       \\
\object{SDSS J133309.19+143706.9} & SDSS\,J1333 & \citet{Szkody+09aj} & 19.11 & 18.48 & 18.15 & 17.94 & 17.90 & 20.95 \\ 
\hline \end{tabular}
\tablefoot{$g_{\rm spec}$ is a spectroscopic apparent magnitude calculated by convolving the SDSS flux-calibrated
spectra with the $g$-band response function. These correspond to a different epoch to the photometric magnitudes,
but may be affected by `slit losses' due to errors in astrometry or fibre positioning.}
\end{table*}

\begin{table*}
\caption{\label{tab:obslog} Log of the observations presented in this work.
The mean magnitudes are calculated using only observations outside eclipses.
\reff{The dates are for the first observation in that observing sequence.
The start and end times indicate the beginning of the first and last exposures in the observing sequence.}}
\centering
\begin{tabular}{lcccccccc} \hline
Target     & \reff{Date of first} & Start time & End time & Telescope/ & Optical &  Number of   & Exposure & Mean      \\
           & \reff{observation} (UT) &  (UT)      &  (UT)    & instrument & element & observations & time (s) & magnitude \\
\hline
SDSS\,J0750 & 2010 02 \reff{24} &      00:27 &      01:10 & NTT\,/\,EFOSC2 & $B_{\rm Tyson}$ filter &  18 &  30--60  & 19.3 \\ 
[2pt]
SDSS\,J0756 & 2010 02 \reff{27} &      00:47 &      03:38 & NTT\,/\,EFOSC2 & $B_{\rm Tyson}$ filter &  65 &    60    & 17.6 \\ 
SDSS\,J0756 & 2010 02 \reff{27} &      05:01 &      05:30 & NTT\,/\,EFOSC2 & $B_{\rm Tyson}$ filter &  13 &    60    & 17.5 \\ 
SDSS\,J0756 & 2010 02 \reff{28} &\reff{00:25}&      03:38 & NTT\,/\,EFOSC2 & $B_{\rm Tyson}$ filter &  96 &    30    & 17.6 \\ 
[2pt]
SDSS\,J0921 & 2009 01 \reff{22} &      05:32 &      09:08 & NTT\,/\,EFOSC2 & $B_{\rm Tyson}$ filter & 162 &  40--80  & 20.5 \\
SDSS\,J0921 & 2009 01 \reff{23} &      03:42 &      05:31 & NTT\,/\,EFOSC2 & $B_{\rm Tyson}$ filter &  89 &    40    & 20.5 \\
SDSS\,J0921 & 2009 01 \reff{23} &      07:37 &      09:06 & NTT\,/\,EFOSC2 & $B_{\rm Tyson}$ filter &  70 &  40--80  & 20.5 \\
SDSS\,J0921 & 2009 01 \reff{27} &      03:33 &      05:00 & NTT\,/\,EFOSC2 & $B_{\rm Tyson}$ filter &  71 &    40    & 20.5 \\
[2pt]
SDSS\,J0924 & 2010 02 \reff{26} &      00:16 &      01:42 & NTT\,/\,EFOSC2 & $B_{\rm Tyson}$ filter &  18 & 180--240 & 19.6 \\ 
[2pt]
SDSS\,J0935 & 2010 02 \reff{24} &      01:19 &      05:07 & NTT\,/\,EFOSC2 & $B_{\rm Tyson}$ filter &  56 &  90--180 & 19.0 \\ 
SDSS\,J0935 & 2010 02 \reff{25} &      04:32 &      05:37 & NTT\,/\,EFOSC2 & $B_{\rm Tyson}$ filter &  21 &  90--120 & 18.8 \\ 
SDSS\,J0935 & 2010 \reff{03 01} &      03:25 &      04:48 & NTT\,/\,EFOSC2 & $B_{\rm Tyson}$ filter &  29 &     90   & 19.0 \\ 
[2pt]
SDSS\,J1006 & 2010 \reff{03 01} &      04:59 &      05:44 & NTT\,/\,EFOSC2 & $B_{\rm Tyson}$ filter &  11 &    180   & 18.5 \\ 
[2pt]
SDSS\,J1057 & 2010 02 \reff{26} &      03:26 &      06:30 & NTT\,/\,EFOSC2 & $B_{\rm Tyson}$ filter &  29 &    300   & 19.5 \\ 
SDSS\,J1057 & 2010 02 \reff{27} &      03:42 &      04:52 & NTT\,/\,EFOSC2 & $B_{\rm Tyson}$ filter &  15 &    210   & 19.5 \\ 
SDSS\,J1057 & 2010 \reff{03 01} &      05:54 &      06:54 & NTT\,/\,EFOSC2 & $B_{\rm Tyson}$ filter &  13 &    210   & 19.4 \\ 
[2pt]
CSS\,J1126  & 2010 02 \reff{26} &      02:07 &      02:51 & NTT\,/\,EFOSC2 & $B_{\rm Tyson}$ filter &  11 &    180   & 18.5 \\ 
CSS\,J1126  & 2010 02 \reff{26} &      06:51 &      07:33 & NTT\,/\,EFOSC2 & $B_{\rm Tyson}$ filter &  13 &    120   & 18.5 \\ 
CSS\,J1126  & 2010 02 \reff{26} &      08:41 &      09:33 & NTT\,/\,EFOSC2 & $B_{\rm Tyson}$ filter &  16 &    120   & 18.4 \\ 
CSS\,J1126  & 2010 02 \reff{27} &      05:34 &      06:10 & NTT\,/\,EFOSC2 & $B_{\rm Tyson}$ filter &  11 &    120   & 18.3 \\ 
CSS\,J1126  & 2010 \reff{03 01} &      07:07 &      08:16 & NTT\,/\,EFOSC2 & $B_{\rm Tyson}$ filter &  21 &    120   & 18.2 \\ 
[2pt]
SDSS\,J1324 & 2009 01 25        &\reff{07:35}&\reff{09:13}& NTT\,/\,EFOSC2 & $B_{\rm Tyson}$ filter &  63 &     60   & 21.1 \\
SDSS\,J1324 & 2009 01 26        &      05:26 &      09:11 & NTT\,/\,EFOSC2 & $B_{\rm Tyson}$ filter & 144 &     60   & 21.1 \\
SDSS\,J1324 & 2009 01 27        &\reff{05:40}&\reff{09:12}& NTT\,/\,EFOSC2 & $B_{\rm Tyson}$ filter & \reff{111} &     80   & 21.1 \\
[2pt]
CSS\,J1325  & 2010 02 \reff{25} &      06:17 &      09:24 & NTT\,/\,EFOSC2 & $B_{\rm Tyson}$ filter &  30 & 240--300 & 19.8 \\ 
CSS\,J1325  & 2010 02 \reff{26} &      07:39 &      08:34 & NTT\,/\,EFOSC2 & $B_{\rm Tyson}$ filter &  11 &    240   & 19.9 \\ 
CSS\,J1325  & 2010 02 \reff{27} &      07:22 &      07:54 & NTT\,/\,EFOSC2 & Grism \#11             &   3 &    900   &      \\ 
CSS\,J1325  & 2010 \reff{03 01} &      08:24 &      09:36 & NTT\,/\,EFOSC2 & $B_{\rm Tyson}$ filter &  14 &    240   & 20.1 \\ 
[2pt]
SDSS\,J1333 & 2010 02 \reff{27} &      06:20 &      06:36 & NTT\,/\,EFOSC2 & $B_{\rm Tyson}$ filter &   4 & 180--240 & 19.7 \\ 
SDSS\,J1333 & 2010 02 \reff{27} &      07:02 &      07:07 & NTT\,/\,EFOSC2 & $B_{\rm Tyson}$ filter &   2 &    240   & 18.9 \\ 
SDSS\,J1333 & 2010 02 \reff{27} &      08:13 &      09:45 & NTT\,/\,EFOSC2 & $B_{\rm Tyson}$ filter &  18 &    240   & 19.3 \\ 
SDSS\,J1333 & 2010 02 \reff{28} &      07:52 &      09:42 & NTT\,/\,EFOSC2 & $B_{\rm Tyson}$ filter &  21 & 210--240 & 19.5 \\ 
SDSS\,J1333 & 2010 05 12        &      23:58 &      02:30 & CA35\,/\,LAICA & $g$ filter             & 145 &     60   & 19.4 \\ 
\hline \end{tabular} \end{table*}

\section{Observations and data reduction}                                                      \label{sec:obs}

The observations presented in this work were obtained using the New Technology Telescope (NTT) at ESO La Silla, equipped with the EFOSC2 focal-reducing instrument%
\footnote{\tt http://www.eso.org/sci/facilities/lasilla/ instruments/efosc/index.html}
\citep{Buzzoni+84msngr}. EFOSC2 was used in imaging mode with a Loral 2048$\times$2048\,px$^2$ CCD, giving a field of view of 4.4$\times$4.4\,arcmin$^2$ at a plate scale of 0.13\apx. For the 2009 observing run we binned the CCD 2$\times$2, but for the 2010 run the CCD was unbinned. All NTT images were obtained with a $B_{\rm Tyson}$ filter (ESO filter \#724), which has a central wavelength of 4445\,\AA\ and a FWHM of 1838\,\AA.

A few additional observations of SDSS\,J1333 were obtained on the night of 2010/05/12, in poor seeing conditions, and using the Calar Alto 3.5\,m telescope and LAICA wide-area camera. The CCD was binned 4$\times$4, yielding an effective plate scale of 1.0\apx, and images of duration 60\,s were obtained through a Gunn $g$ filter.

All data were reduced using the {\sc defot} pipeline, written in {\sc idl}%
\footnote{The acronym {\sc idl} stands for Interactive Data Language and is a trademark of ITT Visual
Information Solutions. For further details see {\tt http://www.ittvis.com/ProductServices/IDL.aspx}.}
 \citep{Me+09mn,Me+14mn}. Aperture photometry was performed using the {\sc astrolib}/{\sc aper} procedure%
\footnote{The {\sc astrolib} subroutine library is distributed by NASA. For further details see {\tt http://idlastro.gsfc.nasa.gov/}.},
which originates from {\sc daophot} \citep{Stetson87pasp}.

The instrumental differential magnitudes were transformed to apparent $V$ magnitudes using formulae from \citet{Jordi++06aa} and SDSS magnitudes of the comparison stars. Given the variety of spectral energy distributions of CVs, and the response function of $B_{\rm Tyson}$ filter we used, the \reff{zeropoints of the} apparent magnitudes are uncertain by at least several tenths of a magnitude.

\subsection{Times of minimum light}

\begin{table}
\caption{\label{tab:eclipses} All available measured
times of eclipse for the objects studied in this work.}
\centering
\begin{tabular}{l r@{\,$\pm$\,}l r r l} \hline
Object & \mc{Time of eclipse}  & Cycle & Residual & Ref. \\
       & \mc{(HJD(UTC))}       &       & (d)      &      \\
\hline
SDSS\,J0750 & 2454853.6293 & 0.0002 &  -277.0 &  0.00016 & 1  \\
SDSS\,J0750 & 2454853.7225 & 0.0002 &  -276.0 &  0.00020 & 1  \\
SDSS\,J0750 & 2454854.5612 & 0.0003 &  -267.0 &  0.00041 & 1  \\
SDSS\,J0750 & 2454857.6353 & 0.0002 &  -234.0 &  0.00005 & 1  \\
SDSS\,J0750 & 2454858.5670 & 0.0003 &  -224.0 &  0.00009 & 1  \\
SDSS\,J0750 & 2454879.4358 & 0.0001 &     0.0 & -0.00017 & 1  \\
SDSS\,J0750 & 2455251.5388 & 0.0001 &  3994.0 &  0.00001 & 2  \\
[2pt]
SDSS\,J0756 & 2455254.5481 & 0.0001 &     0.0 &  0.00000 & 2  \\
SDSS\,J0756 & 2455255.6434 & 0.0001 &     8.0 &  0.00000 & 2  \\
[2pt]
SDSS\,J0924 & 2454856.7192 & 0.0002 &  -250.0 & -0.00013 & 1  \\
SDSS\,J0924 & 2454856.8104 & 0.0003 &  -249.0 & -0.00008 & 1  \\
SDSS\,J0924 & 2454857.7220 & 0.0002 &  -239.0 &  0.00011 & 1  \\
SDSS\,J0924 & 2454858.7245 & 0.0002 &  -228.0 &  0.00006 & 1  \\
SDSS\,J0924 & 2454879.5046 & 0.0002 &     0.0 & -0.00001 & 1  \\
SDSS\,J0924 & 2455253.5476 & 0.0014 &  4104.0 & -0.00001 & 2  \\
[2pt]
SDSS\,J0935 & 2455251.5662 & 0.0004 &     0.0 & -0.00041 & 2  \\
SDSS\,J0935 & 2455251.6309 & 0.0007 &     1.0 &  0.00023 & 2  \\
SDSS\,J0935 & 2455251.6946 & 0.0004 &     2.0 & -0.00013 & 2  \\
SDSS\,J0935 & 2455252.7203 & 0.0004 &    18.0 &  0.00062 & 2  \\
SDSS\,J0935 & 2455256.6912 & 0.0004 &    80.0 & -0.00014 & 2  \\
[2pt]
SDSS\,J1006 & 2454497.6338 & 0.0005 &  -231.0 &  0.00015 & 3  \\
SDSS\,J1006 & 2454540.5802 & 0.0005 &     0.0 &  0.00058 & 3  \\
SDSS\,J1006 & 2454541.5091 & 0.0002 &     5.0 & -0.00009 & 3  \\
SDSS\,J1006 & 2454821.6805 & 0.0002 &  1512.0 & -0.00004 & 3  \\
SDSS\,J1006 & 2455256.7178 & 0.0005 &  3852.0 &  0.00012 & 2  \\
[2pt]
SDSS\,J1057 & 2455253.6903 & 0.0017 &     0.0 & -0.00103 & 2  \\
SDSS\,J1057 & 2455253.7555 & 0.0017 &     1.0 &  0.00136 & 2  \\
SDSS\,J1057 & 2455254.6958 & 0.0017 &    16.0 & -0.00045 & 2  \\
SDSS\,J1057 & 2455256.7690 & 0.0017 &    49.0 &  0.00012 & 2  \\
[2pt]
CSS\,J1126  & 2455253.8126 & 0.0001 &     0.0 & -0.00008 & 2  \\
CSS\,J1126  & 2455253.8905 & 0.0002 &     1.0 &  0.00037 & 2  \\
CSS\,J1126  & 2455254.7420 & 0.0002 &    12.0 & -0.00004 & 2  \\
CSS\,J1126  & 2455256.8331 & 0.0001 &    39.0 &  0.00000 & 2  \\
[2pt]
CSS\,J1325  & 2455252.7705 & 0.0005 &    -1.0 & -0.00034 & 2  \\
CSS\,J1325  & 2455252.8332 & 0.0001 &     0.0 & -0.00002 & 2  \\
CSS\,J1325  & 2455252.8954 & 0.0003 &     1.0 & -0.00019 & 2  \\
CSS\,J1325  & 2455253.8313 & 0.0001 &    16.0 &  0.00007 & 2  \\
CSS\,J1325  & 2455256.8872 & 0.0005 &    65.0 & -0.00044 & 2  \\
\hline \end{tabular}
\tablebib{(1)~\citet{Me+10aa}; (2)~This~work; (3)~\citet{Me+09aa}.}
\end{table}

Eclipse midpoints were measured by shifting each light curve against its own mirror-image until the respective ascending and descending branches were in the best agreement. The time defining the axis of reflection was taken as the midpoint of the eclipse, and uncertainties were estimated based on the shift required for which an offset was obvious. All known times of minimum light for our targets are collected in Table\,\ref{tab:eclipses}.

Due to the poor sky conditions during \reff{the 2010 February} observing run (bright moon and bad seeing) we had to use rather long exposure times to obtain good photometry of our target objects. Some eclipses were therefore covered by only one datapoint, in which case we quote the midpoint of the exposure and take the uncertainty to be half of the exposure time. This poor sampling rate prevents us from using the data to measure the physical properties of the CVs from modelling their light curves.

The results for each system are presented below in three catagories. Firstly we discuss and give the first orbital period measurements of the four new eclipsing CVs. Then we obtain improved ephemerides for four CVs previously known to be eclipsing. Finally, we present light curves obtained for the three magnetic systems SDS\,J0921, SDSS\,J1324 and SDSS\,J1333.


\section{Four new eclipsing cataclysmic variables}                                                \label{sec:eclipsers}

\subsection{SDSS J075653.11$+$085831.8}                                                                \label{sec:0756}

\begin{figure} \includegraphics[width=0.48\textwidth,angle=0]{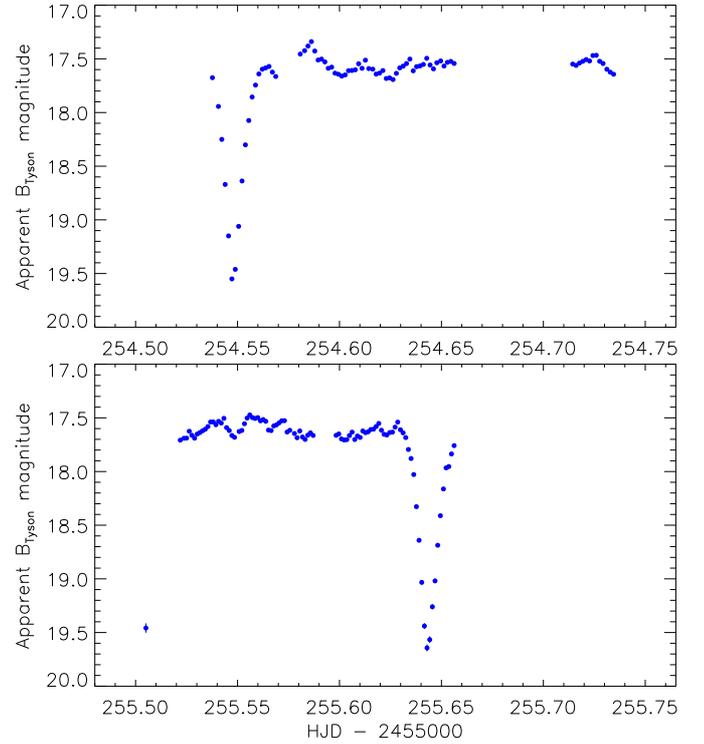} \\
\caption{\label{fig:0756} The light curves of SDSS\,J0756. The errorbars are
mostly smaller than the point sizes.} \end{figure}


SDSS\,J0756 was identified as a variable star in a search for new dwarf novae in existing photometric and astrometric catalogues \citep{Wils+10mn}. Its variability amplitude (1.2\,mag between multiple SDSS photometric observations) was below the 1.5\,mag minimum value used in that work to identify dwarf novae. However, its SDSS spectrum was inspected and found to be typical of the SW\,Sex stars, which are the dominant population of CVs in the 3--4\,hr orbital period interval. A defining characteristic of SW\,Sex stars is spectra which feature a hot continuum with strong He\,II and Bowen blend emission; many of them also show eclipses \citep{Thorstensen+91aj,Rodriguez+07mn2}.

\begin{figure} \includegraphics[width=0.48\textwidth,angle=0]{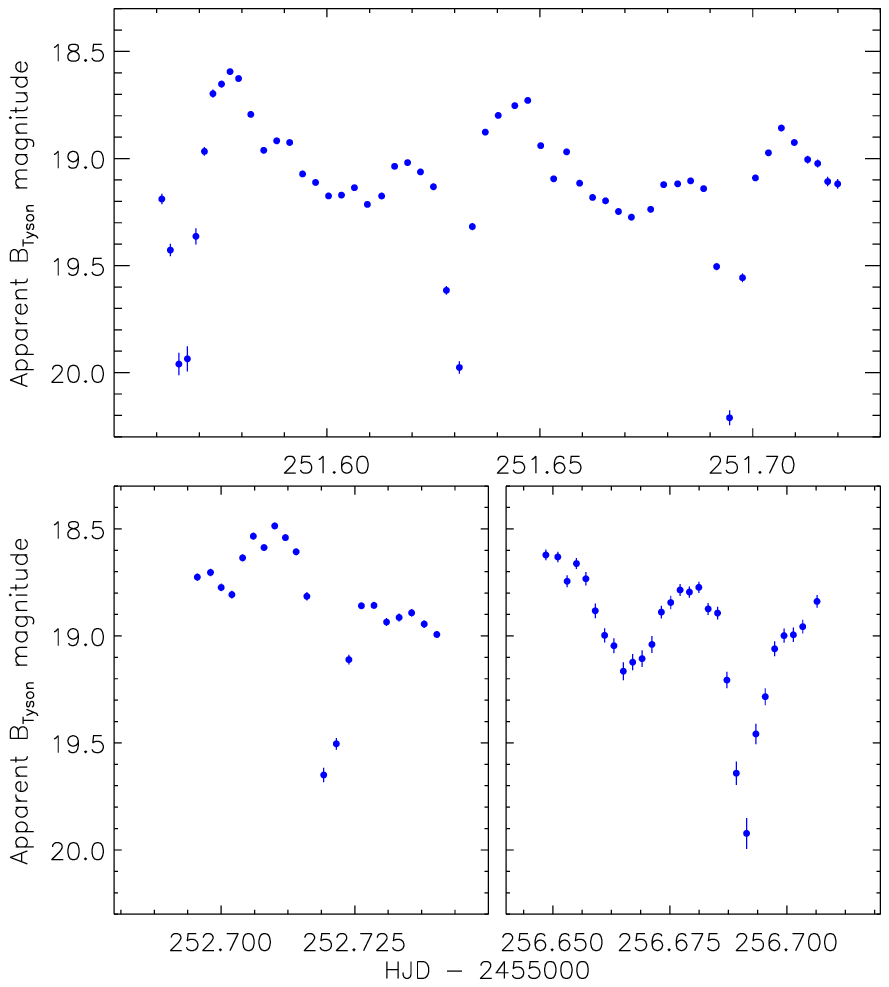} \\
\caption{\label{fig:0935} The light curves of SDSS\,J0935.} \end{figure}

Two complete eclipses of SDSS\,J0756 were observed on successive nights using the NTT (Fig.\,\ref{fig:0756}). On the second of these nights we also obtained a single datapoint which was clearly in eclipse, before an unfortunate gap in observations for telescope maintenance. This single datapoint allows the orbital period of SDSSS\,J0756 to be determined without cycle-count ambiguities, resulting in the ephemeris:
\[ {\rm Min\,I\ (HJD}) = 2455254.5481 (1) + 0.136913 (18) \times E \]
where $E$ is the cycle count and the bracketed numbers represent the uncertainty in the \reff{last digit of the preceding number}. This measurement corresponds to an orbital period of $\Porb = 197.154 \pm 0.025$\,min, which puts SDSS\,J0756 beyond the 2--3\,hr period gap seen in the period distribution of CVs, and right into the 3--4\,hr period interval where most SW\,Sex stars are found. Its light curve is very similar to that of SDSS J075443.01$+$500729.2, an eclipsing SW\,Sex star with a period of 206.0\,min \citep{Me+07mn2}.

The eclipsing nature of SDSS\,J0756 was first announced by \citet{Me++12iaus}. A detailed photometric and spectroscopic study of this object was presented by \citet{Tovmassian+14aj} whilst the current work was nearing completion. Their orbital period measurement is in good agreement with our own.

\subsection{SDSS J093537.46$+$161950.8}                                                                \label{sec:0935}


SDSS\,J0935 was found to be a CV by \citet{Szkody+09aj} due to the presence of Balmer and He\,I emission lines in its SDSS spectrum. He\,II $\lambda$4686 emission is strong, leading Szkody et al.\ to suggest that it may contain a magnetic white dwarf, or alternatively be an old nova. We detected eclipses immediately on pointing the NTT towards it (Fig.\,\ref{fig:0935}). Three consecutive eclipses were seen on the night of \reff{2010/02/24}, followed by two more on subsequent nights. Fitting a straight line to the five times of mid-eclipse (Table\,\ref{tab:eclipses}) yields the orbital ephemeris:
\[ {\rm Min\,I\ (HJD}) = 2455251.56661 (21) + 0.0640591 (53) \times E \]
This measurement corresponds to $\Porb = 92.245 \pm 0.008$\,min: SDSS\,J0935 is a good candidate for follow-up high-speed photometry (see fig.\,2 of \citealt{Littlefair+08mn}).

\begin{figure} \includegraphics[width=0.48\textwidth,angle=0]{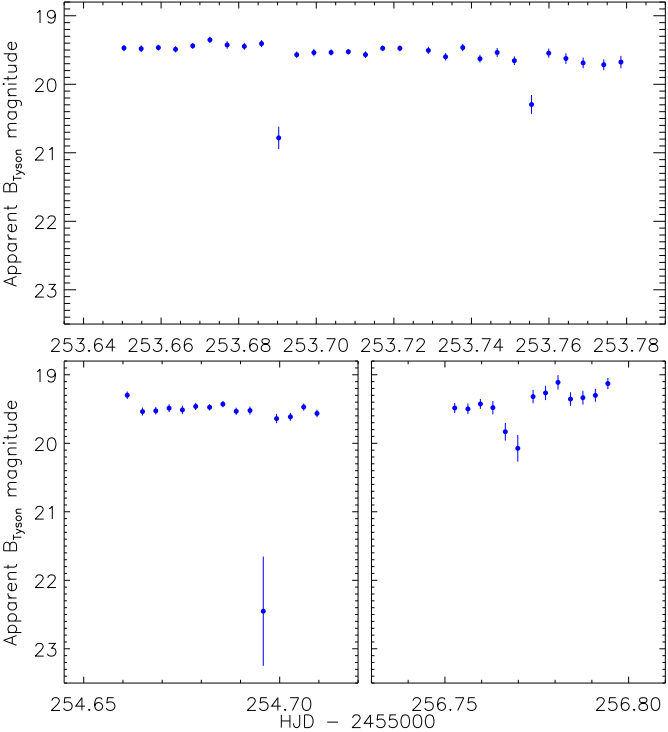} \\
\caption{\label{fig:1057} The light curves of SDSS\,J1057.} \end{figure}

The \reff{2010/02/24} light curve of SDSS\,J0935 presents a notable post-eclipse maximum. Such humps in the light curves of CVs are usually explained as continuum radiation from the impact site of the gas stream from the secondary star on the accretion disc. However, such origin should leave its footprint at orbital phase $\sim$0.8, i.e.\ just before the eclipse. A post-eclipse maximum was observed during outburst for the long-period (0.2096\,d) dwarf nova SDSS J081610.84+453010.2 \citep{Shears+11xxx}, but there the brightness subsequently declines smoothly towards eclipse, leaving the impression of a broad obscuration feature rather than an isolated maximum. \citet{Bailey+88mn} found post-eclipse humps in the magnetic CV WW\,Hor, but in that case they are clearly related to cyclotron emission, and the spectrum of SDSS\,J0935 does not suggest the presence of a strongly magnetic white dwarf. The hump in WW\,Hor is unstable, and changes phase from $\sim$0.2 to $\sim$0.8 between nights.

It therefore appears that post-eclipse hump in SDSS\,J0935 follows a different period to the orbital period, suggesting a relation to superhumps. But these should occur exclusively in superoutburst \reff{for} short period CVs, and the object appears to have been in quiescence during our observations. Still, we note that the spectrum by \citet{Szkody+09aj} presents a rather steep blue continuum and broad, but weak, Balmer emission compared to other short-period dwarf novae. SDSS\,J0935 is certainly an object worthy of further investigation.

\subsection{SDSS J105754.25$+$275947.5}                                                                \label{sec:1057}

\begin{figure} \centering \includegraphics[angle=-90,width=0.48\textwidth]{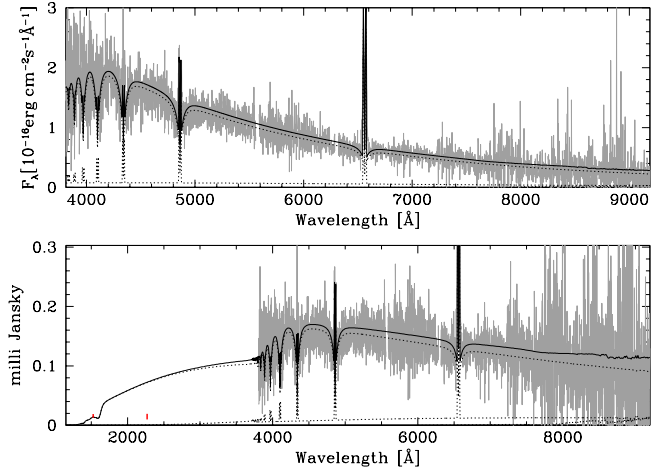} \\
\caption{\label{fig:1057:sed} Three-component model of the SDSS spectrum of SDSS\,J1057. The
three components are a white dwarf with a temperature of 10\,500\,K, and surface gravity of $\log g = 8.0$ (c.g.s.);
an isothermal and isobaric hydrogen slab, and an L5 companion star, all scaled to a distance
of $d = 210$\,pc. The two red points represent data from the GALEX satellite.} \end{figure}


SDSS\,J1057 is another object identified as a CV by \citet{Szkody+09aj}, who suggested that it might be eclipsing due to the double-peaked nature of its H$\alpha$ emission line (Fig.\,\ref{fig:sdssspec}). Here we present the discovery that it is indeed an eclipsing system (Fig.\,\ref{fig:1057}) and the first measurement of its orbital period. We observed two consecutive eclipses on the night of \reff{2010/02/26} and two on later nights. The eclipses are short and deep, and are entirely encompassed by one datapoint in our light curves. We therefore take the midpoints of those datapoints as the times of mid-eclipse, which results in the ephemeris:
\[ {\rm Min\,I (HJD}) = 2455253.6913 (11) + 0.062807 (43) \times E \]
with $\Porb = 90.44 \pm 0.06$\,min. There is an alternative orbital ephemeris with different cycle counts and $\Porb = 96.40 \pm 0.07$\,min, but this can be rejected from inspection of the residuals of the ephemeris fit and of the light curve plotted against orbital phase. Further observations, at a higher time resolution, could provide an independent confirmation of this result.

The light curve of SDSS\,J1057 is rather flat outside eclipse; this is most discernible by comparison to SDSS\,J0935 and CSS\,J1325. Most short-period CVs have a pronounced `orbital hump' immediately before eclipse, caused by the bright spot on the edge of the accretion disc rotating into view. The faintness of the bright spot in SDSS\,J1057 suggests that this system was in a state of very low accretion at the time of our observations. Despite this, there is no sign of the secondary star in the SDSS spectrum (Fig.\,\ref{fig:sdssspec}) even though the white dwarf primary is clearly visible. SDSS\,J1057 is a good candidate for a period-bounce system, and deserves more detailed study.

We have analysed the spectral energy distribution of SDSS\,J1057 using the SDSS spectrum and GALEX fluxes \citep{Morrissey+07apjs}. We find a decent fit to these data using the model of \citet{Gansicke+06mn} with a white dwarf effective temperature of 10\,500\,K, an L5 secondary star, and an accretion disc of temperature 5800\,K, all at a distance of 210\,pc (Fig.\,\ref{fig:1057:sed}). The GALEX near-ultraviolet flux is much lower than predicted, and may have been taken during eclipse. The low white dwarf temperature and late secondary-star spectral type are consistent with SDSS\,J1057 being a post-bounce system with a very low accretion rate.

\subsection{CSS J132536$+$210037}                                                                \label{sec:1325}

\begin{figure} \includegraphics[width=0.48\textwidth,angle=0]{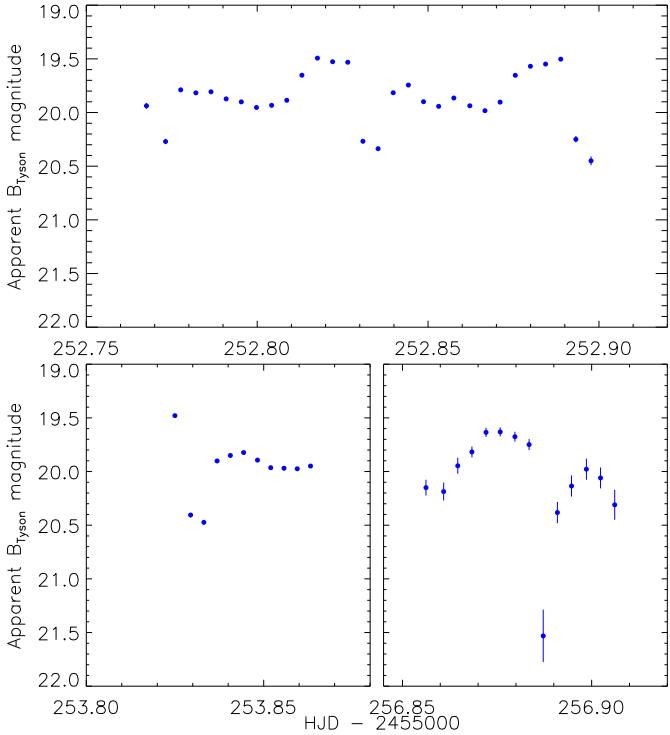} \\
\caption{\label{fig:1325} The light curves of CSS\,J1325.} \end{figure}


CSS\,J1325 was detected as a probable dwarf nova by the {\it Catalina Real Time Transient Survey}\footnote{\tt http://crts.caltech.edu/} (CRTS; \citealt{Drake+09apj}), based on a light curve in which the star was normally around magnitude 20 but twice brightened by at least 1.5\,mag. It was included in a study of dwarf novae by \citet{Wils+10mn}, who also noticed that its SDSS $ugriz$ apparent magnitudes returned highly unusual colour indices. This was interpreted as the possible onset of eclipse during the SDSS photometric observations, which are taken in the order $riuzg$ and with individual integration times of 54.1\,s.

\reff{We therefore targeted CSS\,J1325 as a possible eclipsing CV, obtaining immediate confirmation. The first two brightness measurements we obtained of this object differed by 0.34\,mag.} Three consecutive eclipses were observed on the night of \reff{2010/02/25}, and two more were measured on later nights in our NTT run. We find the orbital ephemeris:
\[ {\rm Min\,I (HJD}) = 2455252.833222 (85) + 0.06223756 (61) \times E \]
which gives $\Porb = 89.821 \pm 0.009$\,min. A spectrum of CSS\,J1325 was not obtained by the SDSS, as its eclipse-affected $ugriz$ apparent magnitudes place it outside the high-priority regions in colour space. However, its eclipsing nature makes CSS\,J1325 well suited to further observations aimed at measuring its physical properties.

\subsubsection{A spectrum of CSS\,J1325}

\begin{figure} \includegraphics[width=0.48\textwidth,angle=0]{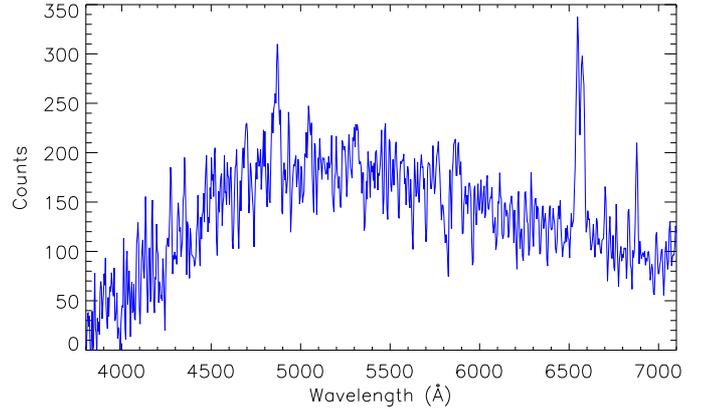} \\
\caption{\label{fig:1325:spec} Spectrum of CSS\,J1325 represented as counts per wavelength
increment. The spectrum has been smoothed with a Savitsky-Golay filter to aid in the
identification of spectral lines.} \end{figure}

Whilst observing we took the opportunity to obtain an identification spectrum of CSS\,J1325 (Fig.\,\ref{fig:1325:spec}). For these observations EFOSC was equipped with grism \#11, yielding a reciprocal dispersion of 2.0\Apx\ and a resolution of 17\,\AA. Three exposures of 900\,s each were obtained, wavelength-calibrated with a helium-argon arc line observation, and combined into one spectrum. Data reduction was performed with the {\sc pamela} and {\sc molly} packages \citep{Marsh89pasp} in the same way as in previous papers in this series \citep{Me+08mn,Me++08mn}.

The final spectrum of CSS\,J1325 (Fig.\,\ref{fig:1325:spec}) is rather noisy, due to the faintness of the target star and the relatively poor seeing, but clearly shows moderately weak emission lines at H$\alpha$ and H$\beta$. The H$\alpha$ emission is double-peaked, a common feature of the spectra of eclipsing CVs (e.g.\ SDSS\,J1057 in Fig.\,\ref{fig:sdssspec}). Based on the above observations, CSS\,J1325 can be classified as a CV which shows both eclipses and dwarf nova outbursts.


\section{New orbital ephemerides for four known eclipsing cataclysmic variables}

\begin{figure} \includegraphics[width=0.48\textwidth,angle=0]{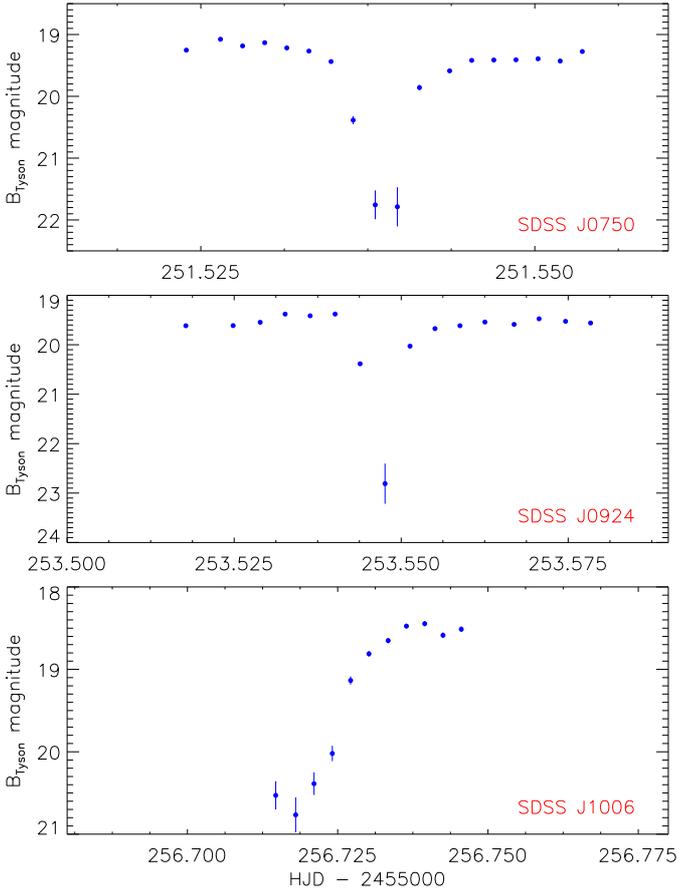} \\
\caption{\label{fig:3ecv} The light curves of the known eclipsing CVs SDSS\,J0750,
SDSS\,J0924 and SDSS\,J1006, obtained in order to improve their orbital ephemeris.}
\end{figure}

\subsection{SDSS J075059.97$+$141150.1}                                                                \label{sec:0750}


SDSS\,J0750 was identified as a CV by \citet{Szkody+07aj}, on the basis of its SDSS spectrum. It was discovered to be eclipsing by \citet{Me+10aa}, who measured an orbital period of $134.1564 \pm 0.0008$\,min. We have observed one further eclipse of this system (Fig.\,\ref{fig:3ecv}), allowing the uncertainty in the orbital period to be lowered by an order of magnitude. We revise the linear orbital ephemeris to:
\[ {\rm Min\,I\ (HJD}) = 2454879.435968 (69) + 0.093165454 (30) \times E \]
corresponding to a period of $134.15825 \pm 0.00004$\,min. The cycle count over the intervening 13 months is unambiguous: the nearest alternative period differs by $45\sigma$ from the original value found by \citet{Me+10aa}.

\subsection{SDSS J092444.48$+$080150.9}                                                                \label{sec:0924}


SDSS\,J0924 \reff{(also named HU\,Leo\footnote{\reff{{\it Simbad} erroneously lists SDSS\,J0924 (HU\,Leo) as a detached eclipsing binary.}})} was identified by \citet{Szkody+05aj} as a possible magnetic CV from an SDSS spectrum which shows strong and narrow Balmer and \ion{He}{II} emission and hints of the secondary star towards the red.  \citet{Me+10aa} found it to be eclipsing, with no obvious sign of an accretion disc, and measured its orbital period to be $131.2432 \pm 0.0014$\,min. We have obtained a light curve covering one additional eclipse (Fig.\,\ref{fig:3ecv}), which was detected in only three datapoints so is very undersampled. We take the midpoint of the middle datapoint as the derived eclipse time, and half the exposure time as its uncertainty. The resulting ephemeris is:
\[ {\rm Min\,I\ (HJD}) = 2454879.50461 (11) + 0.09114108 (31) \times E \]
Cycle count errors can be rejected at the $24\sigma$ level.

\begin{figure} \includegraphics[width=0.48\textwidth,angle=0]{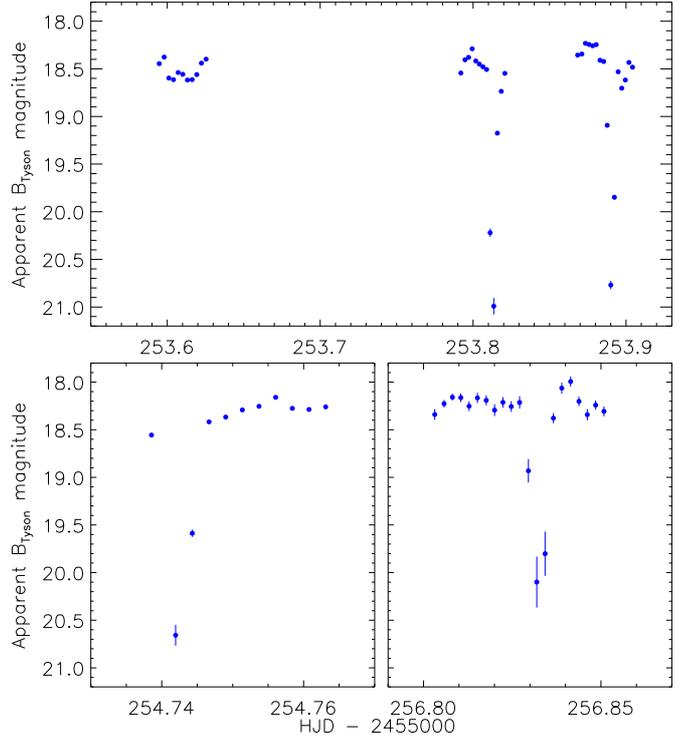} \\
\caption{\label{fig:1126} The light curves of CSS\,J1126.} \end{figure}

\subsection{SDSS J100658.40$+$233724.4}                                                                \label{sec:1006}


SDSS\,J1006 was originally identified as a CV by \citet{Szkody+07aj}, from an SDSS spectrum showing strong and wide Balmer emission lines. \citet{Me+09aa} obtained extensive photometry and spectroscopy, from which they measured the orbital period of the system ($267.71507 \pm 0.00060$\,min) and the masses and radii of the component stars. We obtained a light curve of part of one additional eclipse (Fig.\,\ref{fig:3ecv}), doubling the temporal coverage of the photometric observations of this system. We find an improved orbital ephemeris of:
\[ {\rm Min\,I (HJD}) = 2454540.57962 (16) + 0.18591331 (12) \times E \]
SDSS\,J1006 is a long-period CV with $\Porb = 267.71516 \pm 0.00017$\,min.

\subsection{CSS J112634$-$100210}                                                                       \label{sec:css}



CSS\,J1126 was identified as an eclipsing CV from photometric observations taken by the CRTS. A dwarf nova outburst of amplitude 3\,mag is also noticable in these data \citep{Drake+08atel}. A subsequent spectrum confirmed the CV classification and revealed a `blue continuum with numerous H and He lines in emission' \citep{Djorgovski+08atel}. CSS\,J1126 is positioned in an area of sky which was not covered in the SDSS spectroscopic observations, so does not have an SDSS spectrum.

We observed this object with the NTT in order to confirm its eclipsing nature and provide the first measurement of its orbital period. Two eclipses were observed in three short light curves taken on the night of \reff{2010/02/26}, and two more eclipses were targeted on subsequent nights. From the measured times of mid-eclipse we find the orbital ephemeris:
\[ {\rm Min\,I (HJD}) = 2455253.812683 (86) + 0.0774466 (34) \times E \]
which corresponds to $\Porb = 111.523 \pm 0.005$\,min. The light curve of CSS\,J1126 is similar to that of SDSS\,J1057, in that it is fairly flat outside eclipse. This implies that either the accretion rate was low or the bright spot was optically thick at the time of our observations.


\section{Three magnetic cataclysmic variables}                                                          \label{sec:mag}

\subsection{SDSS J092122.84$+$203857.1}                                                                \label{sec:0921}

\begin{figure} \includegraphics[width=0.48\textwidth,angle=0]{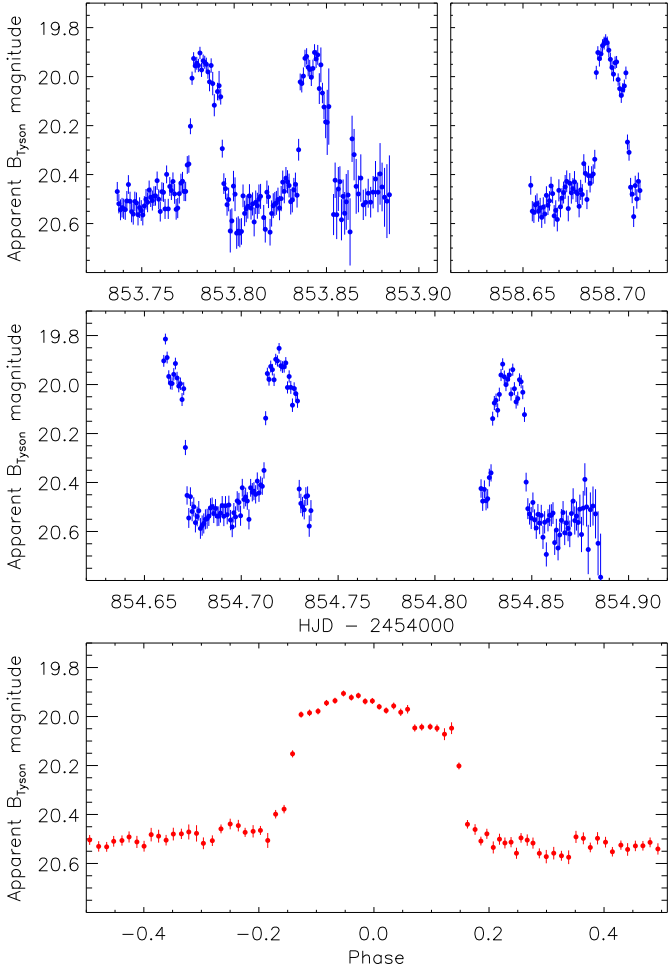} \\
\caption{\label{fig:0921} The light curves of SDSS\,J0921 (upper panels). The
lowest panel shows the photometric measurements as a function of orbital phase
and binned by a factor of 5.} \end{figure}

\begin{table}
\caption{\label{tab:0921} Times of mid-brightening
measured from our light curves of SDSS\,J0921.}
\centering
\begin{tabular}{l r@{\,$\pm$\,}l r r} \hline
Object  &  \mc{Time of mid-brightening}   & Cycle & Residual  \\
        &  \mc{(HJD $-$ 2400000}  &       & (d)       \\
\hline
SDSS\,J0921 & 54853.7847 & 0.0002 &  0.0 & $-$0.00014 \\
SDSS\,J0921 & 54853.8435 & 0.0004 &  1.0 &    0.00016 \\
SDSS\,J0921 & 54854.7208 & 0.0002 & 16.0 & $-$0.00004 \\
SDSS\,J0921 & 54854.8380 & 0.0002 & 18.0 &    0.00016 \\
SDSS\,J0921 & 54858.6988 & 0.0002 & 84.0 & $-$0.00003 \\
\hline \end{tabular}
\end{table}

SDSS\,J0921 was discovered to be a magnetic CV of polar type by \citet{Schmidt+08pasp}. Its SDSS spectrum (Fig.\,\ref{fig:sdssspec}) contains four clear cyclotron humps, which indicate that the white dwarf has a magnetic field strength of 32\,MG. The follow-up spectropolarimetric observations obtained by \citet{Schmidt+08pasp} show variability in both polarisation and flux distribution. Schmidt et al.\ interpreted their observations as evidence of a positively polarised accretion region visible at all times, plus a negatively polarised region visible for only a small fraction of each orbital period. The orbital period of the system was only constrained to be greater than approximately 1.5\,hr.

We observed SDSS\,J0921 on three nights in 2009 January, at which time it displayed brightenings of 0.6\,mag amplitude, occurring every 84\,min and lasting roughly 30\,min (Fig.\,\ref{fig:0921}). These brightenings are telltale signs of an accretion column near the surface of the magnetic white dwarf rotating into and out of view. In order to find the orbital ephemeris we determined the mid-points of these brightenings in exactly the same way as eclipses were measured for the systems above. The mid-points are given in Table\,\ref{tab:0921}, and result in the ephemeris:
\[ {\rm Max (HJD}) = 2454853.78484 (13) + 0.0584999 (30) \times E \]
Under the reasonable assumption that these brightening represent the orbital period of the system, we find $\Porb = 84.240 \pm 0.004$\,min.

The photometric variations of SDSS\,J0921 are strikingly similar to those of EU\,Cnc \citep{Gilliland+91aj,Nair+05ibvs} and VV\,Pup \citep{WarnerNather72}. These two objects are AM\,Her-type magnetic CVs, with orbital periods of 125.5 and 100.4 min, respectively. The variability in their light curves is thought to be due to cyclotron emission from accretion columns above the magnetic poles of the white dwarf. In the case of VV\,Pup, spectroscopic observations have verified that the photometric period coincides with the orbital period \citep[e.g][]{SchneiderYoung80apj}, supporting our assertion above. Such observations have not been secured for EU\,Cnc.

\subsection{SDSS J132411.57$+$032050.4}                                                                \label{sec:1324}

\begin{figure} \includegraphics[width=0.48\textwidth,angle=0]{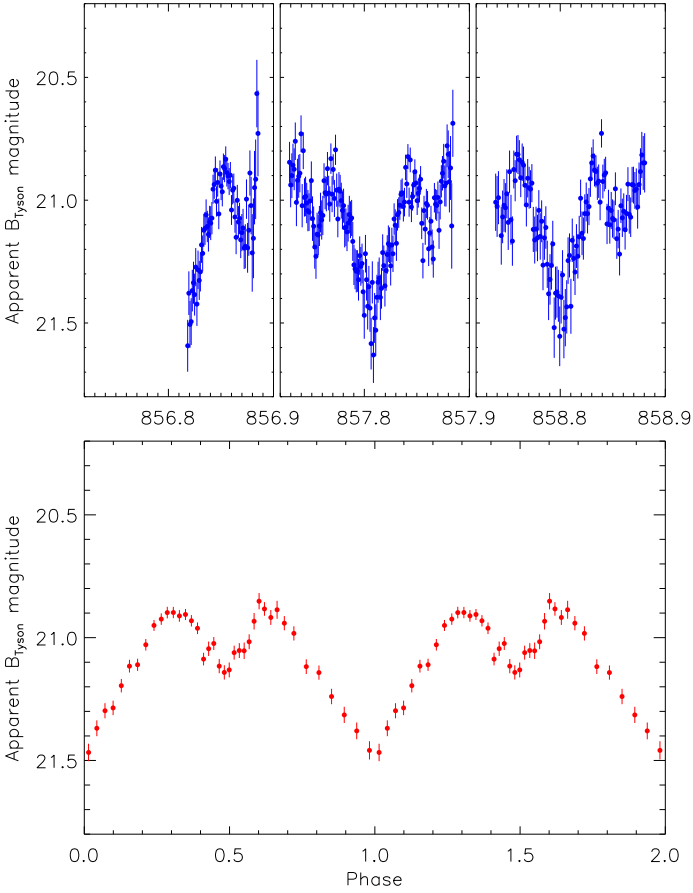} \\
\caption{\label{fig:1324} The light curves of SDSS\,J1324 (upper panels). The
lowest panel shows the photometric measurements as a function of phase and
binned by a factor of 8. Two phases are reproduced for display purposes.} \end{figure}

SDSS\,J1324 (also named PZ\,Vir) is a faint magnetic CV discovered by \citet{Szkody+03apj} from its SDSS spectrum, which shows a very faint object ($g = 23.3$) with a large flux excess around 5600\,\AA. From this and two other fainter cyclotron features, \citet{Szkody+03apj} inferred that the white dwarf has a magnetic field strength of 63\,MG. They also obtained a short light curve which showed variability at a period of roughly 2.6\,hr, and spectropolarimetry which demonstrated that the cyclotron feature is highly circularly polarised. SDSS\,J1324 has a very low accretion rate in which the accretion energy is not dissipated in a shock but instead is efficiently converted into optical cyclotron emission at and below the surface of the white dwarf (labelled the `bombardment scenario'). \citet{Szkody+04aj} obtained an XMM-Newton observation of the system which showed it to be a very weak X-ray source, in agreement with this scenario. \citet{Schmidt+05apj} presented spectropolarimetry of SDSS\,J1324 from which they measured a spectroscopic orbital period of $157 \pm 14$\,min and deduced that the spin frequency of the white dwarf is locked to the orbital frequency.

We observed SDSS\,J1324 on three consecutive nights in 2009 January, at which time it was showing clear periodic variability (Fig.\,\ref{fig:1324}). Periodograms were calculated from the light curves using the \citet{Scargle82apj} method, analysis of variance \citep{Schwarzenberg89mn} and orthogonal polynomial \citep{Schwarzenberg96apj} approaches, as implemented within the {\sc tsa}\footnote{\tt http://www.eso.org/projects/esomidas/doc/user/ 98NOV/volb/node220.html} context in {\sc midas}. In all cases the best period was in the region of 159\,min, and other peaks in the periodograms led to phased light curves with a much larger scatter. Taking into account the range of results found using the different periodogram methods, we arrive at a final period measurement of $158.72 \pm 0.10$\,min.

The light curve phased using this period measurement is plotted in the lower panel of Fig.\,\ref{fig:1324}. Its morphology is notably reminiscent of the light curve of the prototypical polar AM\,Her presented by \citet{SzkodyBrownlee77apj}. The variability is likely due to different degrees of cyclotron beaming towards Earth as the angle between the line of sight and the magnetic field axis changes over the orbit \citep{Gansicke+01aa}. \citet{Ferrario++05aspc} have studied the existing spectropolarimetric observations of SDSS\,J1324 and found good results with a model where the accretion energy is released relatively deep inside the white dwarf. The much improved accuracy of our period measurement for SDSS\,J1324 will help in the understanding of this candidate low-accretion-rate polar system \citep{Schmidt+05apj}.

%
%
%
%

\subsection{SDSS J133309.19$+$143706.9}                                                                \label{sec:1333}

\begin{figure} \includegraphics[width=0.48\textwidth,angle=0]{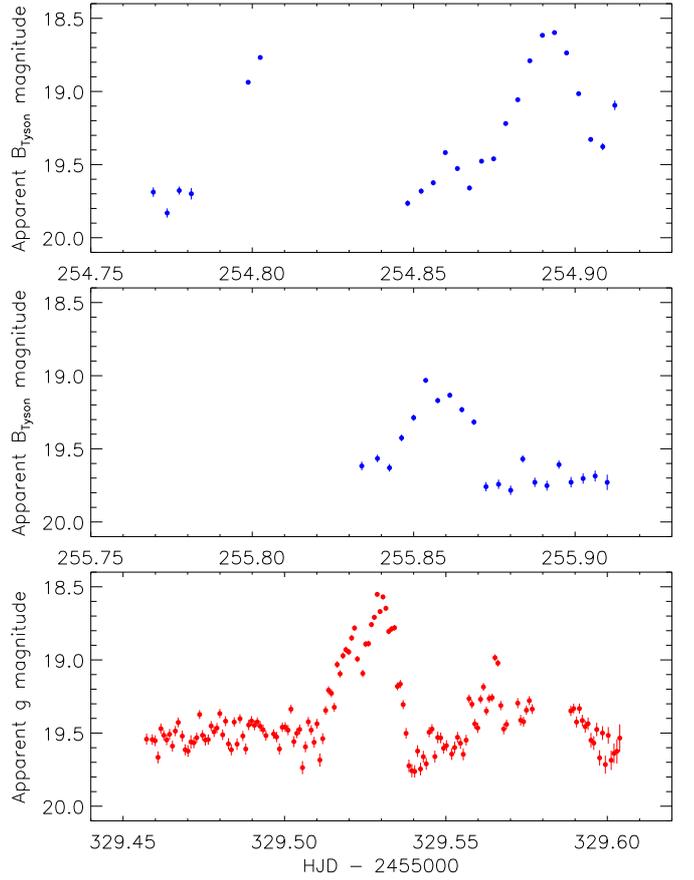} \\
\caption{\label{fig:1333} Light curves of SDSS\,J1333. The upper two panels
show data from the NTT and the lower panel data from the Calar Alto 3.5\,m
telescope.} \end{figure}

SDSS\,J1333 was found to be a magnetic CV by \citet{Schmidt+08pasp}, who measured an orbital period of $132 \pm 6$\,min from spectroscopic velocity measurements of its H$\alpha$ emission line. Our observations, obtained from two telescopes, show obvious variations in the optical brightness of this system (Fig.\,\ref{fig:1333}). Our data are consistent with an orbital period of 132\,min but are insufficient to improve on this value. The light curve shape is reminiscent of the cyclotron-beaming brightenings displayed by SDSS\,J0921, suggesting that the orbital period of SDSS\,J1333 would be relatively easy to obtain from photometric observations with good coverage of all orbital phases.


\section{The orbital period distribution of eclipsing CVs}

\begin{figure} \includegraphics[width=0.48\textwidth,angle=0]{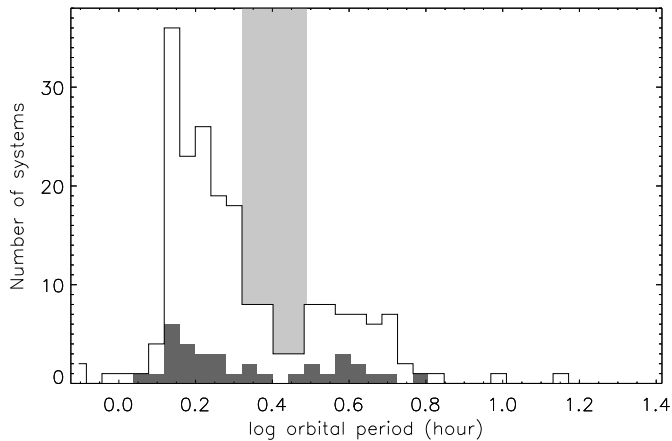} \\
\caption{\label{fig:pd:sdss} The orbital period distribution of CVs identified by the
SDSS (white histogram) and of the subset of these which are eclipsing (grey histogram).
The light grey rectangle delineates the period gap at 2.1--3.1 hours. The periods have
been collected into histogram bins which are of equal size in log space.} \end{figure}

\begin{figure} \includegraphics[width=0.48\textwidth,angle=0]{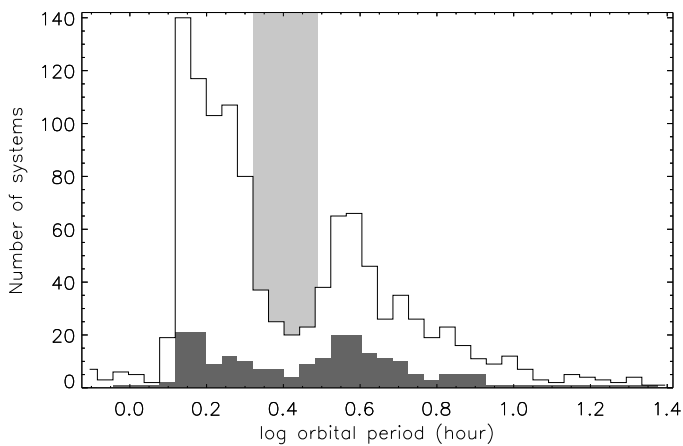} \\
\caption{\label{fig:pd:rk} As for Fig.\,\ref{fig:pd:sdss} but for the RKCat CVs.} \end{figure}

The period distribution of CVs is an important observable quantity for comparison with theoretical population synthesis models \citep{Patterson98pasp,Knigge06mn,Gansicke+09mn}. CVs evolve from longer to shorter orbital periods through the loss of orbital angular momentum, before reaching a minimum period caused by changes in the structure of the secondary star and `bouncing' back to longer periods. However, the observed orbital period distribution of CVs has persistently failed to match theoretical results \citep{Downes+01pasp,RitterKolb03aa}, which predict a large accumulation of objects at the minimum period due to the long evolutionary timescale there. \citet{Gansicke+09mn} identified this `period spike' for the first time, using the observed period distribution of the SDSS CVs. These authors demonstrated that the marked deficiency of short-period CVs could be an observational bias as these systems are intrinsically much fainter than longer-period systems.

Selection effects arise from the limiting magnitude and the method used to detect CVs \citep{Gansicke+09mn}: identification via a blue colour (e.g.\ the Palomar-Green survey; \citealt{Green++86apjs}) or low-resolution survey spectra (e.g.\ the Hamburg Quasar Survey; \citealt{Hagen+95aas}) tends to yield objects with a high accretion luminosity which are predominantly of longer orbital period. Medium-resolution survey spectroscopy (e.g.\ SDSS) yields samples of CVs which are comparatively unbiased, whereas CV discovery via outbursts is biased towards shorter-period objects (\citealt{Uemura10pasj,ThorstensenSkinner12aj,Woudt+12mn}; see also \citealt{Drake+14mn}).

Selection effects also arise from the observational methods used to measure orbital periods. CVs do not give up their secrets easily, especially those which have high accretion rates. This leads to a bias against longer-period systems with higher accretion rates, as their periods are more difficult to measure and observers give them a lower priority for the same reason.

\begin{figure} \includegraphics[width=0.48\textwidth,angle=0]{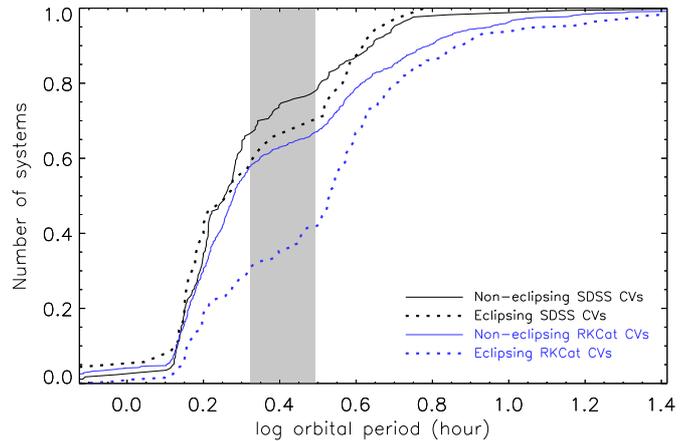} \\
\caption{\label{fig:pd:cumu} Cumulative distribution of the orbital periods of SDSS and RKCat CVs,
both eclipsing and not eclipsing. The 2.1--3.1\,hour period gap is shaded light grey.} \end{figure}

\begin{figure} \includegraphics[width=0.48\textwidth,angle=0]{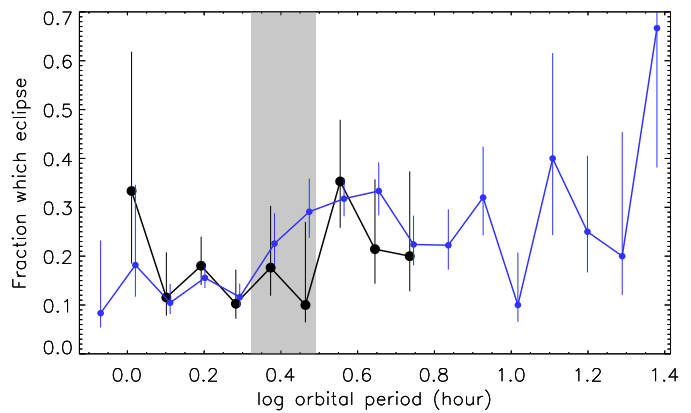} \\
\caption{\label{fig:pd:ecfrac} The fraction of CVs which are eclipsing in the
SDSS sample (black bold symbols) and in RKCat (blue symbols). The samples have
been combined into 17 bins for display purposes and 68.3\% confidence intervals
from binomial statistics are plotted. The RKCat points are shifted by +0.01 in
the abscissa to make them more visible} \end{figure}

The study of eclipses is one of most straightforward ways to measure a CV orbital period, so has an important part to play in determining their period distribution. The likelihood of eclipses is a relatively flat function of orbital period \citep{Warner95book}, but there are observational selection effects \ref{both in} favour of shorter periods (less telescope time is required per object) and against shorter periods (they are intrinsically fainter objects). Forthcoming large-scale sky surveys which are aimed at characterising the faint variable sky (e.g.\ LSST; \citealt{Ivesic+08xxx}) will identify a large number of CVs. Most will be too faint for spectroscopic study with current facilities, so the investigation of these objects will rely heavily on the eclipsing ones.

Fig.\,\ref{fig:pd:sdss} shows the orbital period distribution of all CVs identified from SDSS observations (data taken from \citealt{Gansicke+09mn} with updates). The prevalence of shorter-period systems is clear, and the fraction which are known to eclipse shows no significant trend with orbital period. Fig.\,\ref{fig:pd:rk} shows the known population of CVs according to version 7.20 (July 2013) of the \citet{RitterKolb03aa} catalogue (hereafter RKCat); note that this includes the SDSS CVs. A greater fraction of these objects have periods longer than the 2.1--3.1\,hr period gap.

Fig.\,\ref{fig:pd:cumu} represents these results as cumulative distributions, plotted for eclipsing and non-eclipsing CVs from the SDSS and the RKCat sample. A slightly higher fraction of the SDSS CVs are shortward of the period gap, and the two distributions are similar to each other and to that of the RKCat non-eclipsing CVs. The eclipsing RKCat CVs, however, are predominantly longer-period: this is the only one of the four samples for which the majority of CVs are longward of the period gap. Possible explanations of this include that the eclipses in a significant number of the short-period RKCat CVs have so far evaded detection or are not flagged as eclipsing in RKCat, and that orbital period measurements in non-eclipsing long-period CVs are difficult so these objects are under-represented in RKCat.

Fig.\,\ref{fig:pd:ecfrac} shows the fraction of the SDSS and RKCat samples which eclipse, as a function of orbital period. Few conclusions can be drawn for the SDSS sample, due to small-number statistics, but it is apparent that the eclipsing fraction does not have a strong dependence on period. The RKCat sample, however, shows that a greater fraction of known longer-period CVs exhibit eclipses.


\section{Summary}

We present NTT\,/\,EFOSC2 time-series photometry of eleven CVs, eight of which show eclipses and three of which accomodate magnetic white dwarfs. Four of the targets were not previously known to be eclipsing, and for these plus a fifth object we provide the first measurement of their orbital periods (Table\,\ref{tab:result}). These CVs are prime candidates for detailed follow-up observations of their eclipses, from which their physical properties can be measured to high precision.

The newly-discovered eclipsing systems include CSS\,J1325, which has an orbital period of 89.921\,min, SDSS\,J1057 (90.44\,min), SDSS\,J0935 (92.245\,min) and CSS\,J1126 (111.523\,min). We confirm the eclipsing nature of SDSS\,J0756 (197.154\,min), which was recently discovered by \citet{Tovmassian+14aj}. Along with SDSS\,J0750 (134.158\,min) and the seven systems considered in \citet{Littlefair+08mn}, they form a sequence which would allow the physical properties of short-period CVs to be empirically defined as a function of orbital period. We present the first spectrum of CSS\,J1325, which confirms its classification as an accreting binary system.

\begin{table} \begin{center}
\caption{\label{tab:result} Summary of the orbital periods and CV
classification obtained for the objects studied in this work.}
\begin{tabular}{l c l} \hline
Object      & Period (min)          & Notes                      \\
\hline
SDSS\,J0750 & $134.15825 \pm 0.00004$ & Known eclipsing CV \\
SDSS\,J0756 & $197.154 \pm 0.025$     & Known eclipsing CV \\
SDSS\,J0921 & $84.240 \pm 0.004$      & Magnetic CV \\
SDSS\,J0924 & $131.2432 \pm 0.0004$   & Known eclipsing CV \\
SDSS\,J0935 & $92.245 \pm 0.008$      & New eclipsing CV \\
SDSS\,J1006 & $267.71516 \pm 0.00017$ & Known eclipsing CV \\
SDSS\,J1057 & $90.44 \pm 0.06$        & New eclipsing CV \\
CSS\,J1126  & $111.523 \pm 0.005$     & First period measurement \\
SDSS\,J1324 & $158.72 \pm 0.10$       & Magnetic CV \\
CSS\,J1325  & $89.821 \pm 0.009$      & New eclipsing dwarf nova \\
SDSS\,J1333 & ($132 \pm 6$)           & Magnetic CV \\
\hline \end{tabular} \end{center} \end{table}

We also present photometry of three magnetic CVs and measure orbital periods for two of these, SDSS\,J0921 (84.240\,min) and SDSS\,J1324 (158.72\,min). The spectra and light curves of both are dominated by polarised emission arising from cyclotron radiation, and are strongly reminiscent of the magnetic CVs EU\,Cnc and AM\,Her, respectively. Finally, light curves of the magnetic system SDSS\,J1333 were obtained which confirm its optical variability but do not yield a unique period measurement.

Five of the seven CVs for which we present the first orbital period measurements have periods shorter than the 2--3\,hr gap observed in the general CV population \citep{WhyteEggleton80mn}. This is in line with previous results from our survey of SDSS CVs, where the general population of CVs is dominated by faint short-period systems \citep{Gansicke+09mn}. The remaining two CVs are representatives of the AM\,Her and SW\,Sex classes of accreting binary systems, and have orbital periods within and beyond the period gap, respectively.

We construct the orbital period distributions of all SDSS CVs and of eclipsing SDSS CVs, finding that the fraction of eclipsing objects is not stongly dependent on orbital period. We perform this analysis for all CVs with a known orbital period in RKCat. Whilst the fraction of eclipsing systems is comparable to the SDSS sample at shorter orbital periods, it rises towards longer periods. RKCat is deficient in eclipsing CVs shortward of the period gap. The orbital period distribution of eclipsing CVs will be a key tracer of the population characteristics of CVs discovered in the future by deep sky surveys such as the LSST.


\begin{acknowledgements}

The reduced observational data presented in this work will be made available at the CDS ({\tt http://cdsweb.u-strasbg.fr/}) and at {\tt http://www.astro.keele.ac.uk/$\sim$jkt/}. We are grateful to Stuart Littlefair for suggesting CSS\,J1126 as a worthwhile target. JS acknowledges support from STFC in the form of an Advanced Fellowship. CMC and BTG acknowledge financial support from STFC in the form of grant number ST/F002599/1. The research leading to these results has received funding from the European Research Council under the European Union's Seventh Framework Programme (FP/2007-2013) / ERC Grant Agreement n.\,320964 (WDTracer). BTG was supported in part by the UK’s Science and Technology Facilities Council (ST/I001719/1) Based on observations made with ESO Telescopes at the La Silla Observatory under programme ID 084.D-0056. The following internet-based resources were used in research for this paper: the ESO Digitized Sky Survey; the NASA Astrophysics Data System; the SIMBAD database operated at CDS, Strasbourg, France; the ar$\chi$iv scientific paper preprint service operated by Cornell University; and the AAVSO Variable Star Index.

\end{acknowledgements}



\begin{thebibliography}{66}
\expandafter\ifx\csname natexlab\endcsname\relax\def\natexlab#1{#1}\fi

\bibitem[{{Araujo-Betancor} {et~al.}(2005){Araujo-Betancor}, {G{\"a}nsicke},
  {Long}, {Beuermann}, {de Martino}, {Sion}, \& {Szkody}}]{Araujo+05apj}
{Araujo-Betancor}, S., {G{\"a}nsicke}, B.~T., {Long}, K.~S., {et~al.} 2005,
  ApJ, 622, 589

\bibitem[{{Bailey} {et~al.}(1988){Bailey}, {Wickramasinghe}, {Hough}, \&
  {Cropper}}]{Bailey+88mn}
{Bailey}, J., {Wickramasinghe}, D.~T., {Hough}, J.~H., \& {Cropper}, M. 1988,
  MNRAS, 234, 19P

\bibitem[{{Buzzoni} {et~al.}(1984){Buzzoni}, {Delabre}, {Dekker}, {D'Odorico},
  {Enard}, {Focardi}, {Gustafsson}, {Nees}, {Paureau}, \&
  {Reiss}}]{Buzzoni+84msngr}
{Buzzoni}, B., {Delabre}, B., {Dekker}, H., {et~al.} 1984, The Messenger, 38, 9

\bibitem[{{Dillon} {et~al.}(2008){Dillon}, {G{\"a}nsicke}, {Aungwerojwit},
  {Rodr{\'{\i}}guez-Gil}, {Marsh}, {Barros}, {Szkody}, {Brady}, {Krajci}, \&
  {Oksanen}}]{Dillon+08mn}
{Dillon}, M., {G{\"a}nsicke}, B.~T., {Aungwerojwit}, A., {et~al.} 2008, MNRAS,
  386, 1568

\bibitem[{{Djorgovski} {et~al.}(2008){Djorgovski}, {Glikman}, {Mahabal},
  {Donalek}, {Drake}, {Williams}, {Graham}, {Nugent}, {Thomas}, {Hennawi},
  {Beshore}, {Larson}, \& {Christensen}}]{Djorgovski+08atel}
{Djorgovski}, S.~G.~., {Glikman}, E., {Mahabal}, A., {et~al.} 2008, ATel, 1416

\bibitem[{{Downes} {et~al.}(2001){Downes}, {Webbink}, {Shara}, {Ritter},
  {Kolb}, \& {Duerbeck}}]{Downes+01pasp}
{Downes}, R.~A., {Webbink}, R.~F., {Shara}, M.~M., {et~al.} 2001, PASP, 113,
  764

\bibitem[{{Drake} {et~al.}(2009){Drake}, {Djorgovski}, {Mahabal}, {Beshore},
  {Larson}, {Graham}, {Williams}, {Christensen}, {Catelan}, {Boattini},
  {Gibbs}, {Hill}, \& {Kowalski}}]{Drake+09apj}
{Drake}, A.~J., {Djorgovski}, S.~G., {Mahabal}, A., {et~al.} 2009, ApJ, 696,
  870

\bibitem[{{Drake} {et~al.}(2014){Drake}, {Gaensicke}, {Djorgovski}, {Wils},
  {Mahabal}, {Graham}, {Yang}, {Williams}, {Catelan}, {Prieto}, {Donalek},
  {Larson}, \& {Christensen}}]{Drake+14mn}
{Drake}, A.~J., {Gaensicke}, B.~T., {Djorgovski}, S.~G., {et~al.} 2014, MNRAS,
  in press, {\tt arXiv:1404.3732}

\bibitem[{{Drake} {et~al.}(2008){Drake}, {Williams}, {Mahabal}, {Djorgovski},
  {Graham}, {Beshore}, {Larson}, \& {Christensen}}]{Drake+08atel}
{Drake}, A.~J., {Williams}, R., {Mahabal}, A., {et~al.} 2008, ATel, 1399

\bibitem[{{Ferrario} {et~al.}(2005){Ferrario}, {Wickramasinghe}, \&
  {Schmidt}}]{Ferrario++05aspc}
{Ferrario}, L., {Wickramasinghe}, D.~T., \& {Schmidt}, G.~D. 2005, in
  Astronomical Society of the Pacific Conference Series, Vol. 330, The
  Astrophysics of Cataclysmic Variables and Related Objects, ed. {J.-M.~Hameury
  \& J.-P.~Lasota}, 411--412

\bibitem[{{G{\"a}nsicke} {et~al.}(2009){G{\"a}nsicke}, {Dillon}, {Southworth},
  {Thorstensen}, {Rodr{\'{\i}}guez-Gil}, {Aungwerojwit}, {Marsh}, {Szkody},
  {Barros}, {Casares}, \& {Others}}]{Gansicke+09mn}
{G{\"a}nsicke}, B.~T., {Dillon}, M., {Southworth}, J., {et~al.} 2009, MNRAS,
  397, 2170

\bibitem[{{G{\"a}nsicke} {et~al.}(2001){G{\"a}nsicke}, {Fischer}, {Silvotti},
  \& {de Martino}}]{Gansicke+01aa}
{G{\"a}nsicke}, B.~T., {Fischer}, A., {Silvotti}, R., \& {de Martino}, D. 2001,
  A\&A, 372, 557

\bibitem[{{G{\"a}nsicke} {et~al.}(2006){G{\"a}nsicke}, {Rodr{\'{\i}}guez-Gil},
  {Marsh}, {de Martino}, {Nestoras}, {Szkody}, {Aungwerojwit}, {Barros},
  {Dillon}, {Araujo-Betancor}, \& {Others}}]{Gansicke+06mn}
{G{\"a}nsicke}, B.~T., {Rodr{\'{\i}}guez-Gil}, P., {Marsh}, T.~R., {et~al.}
  2006, MNRAS, 365, 969

\bibitem[{{Gilliland} {et~al.}(1991){Gilliland}, {Brown}, {Duncan}, {Suntzeff},
  {Lockwood}, {Thompson}, {Schild}, {Jeffrey}, \& {Penprase}}]{Gilliland+91aj}
{Gilliland}, R.~L., {Brown}, T.~M., {Duncan}, D.~K., {et~al.} 1991, AJ, 101,
  541

\bibitem[{{Green} {et~al.}(1986){Green}, {Schmidt}, \&
  {Liebert}}]{Green++86apjs}
{Green}, R.~F., {Schmidt}, M., \& {Liebert}, J. 1986, ApJS, 61, 305

\bibitem[{{Hagen} {et~al.}(1995){Hagen}, {Groote}, {Engels}, \&
  {Reimers}}]{Hagen+95aas}
{Hagen}, H.-J., {Groote}, D., {Engels}, D., \& {Reimers}, D. 1995, A\&AS, 111,
  195

\bibitem[{{Hellier}(2001)}]{Hellier01book}
{Hellier}, C. 2001, {Cataclysmic Variable Stars: How and Why they Vary}
  (Springer-Praxis books in astronomy and space science, Springer Verlag, New
  York)

\bibitem[{{Horne} {et~al.}(1991){Horne}, {Wood}, \& {Stiening}}]{Horne++91apj}
{Horne}, K., {Wood}, J.~H., \& {Stiening}, R.~F. 1991, ApJ, 378, 271

\bibitem[{{Ivezi{\'c}} {et~al.}(2008){Ivezi{\'c}}, {Tyson}, {Acosta}, \&
  {Others}}]{Ivesic+08xxx}
{Ivezi{\'c}}, {\v{Z}}., {Tyson}, J.~A., {Acosta}, E., \& {Others}, M. 2008,
  {\tt arXiv:0805.2366}

\bibitem[{{Jordi} {et~al.}(2006){Jordi}, {Grebel}, \& {Ammon}}]{Jordi++06aa}
{Jordi}, K., {Grebel}, E.~K., \& {Ammon}, K. 2006, A\&A, 460, 339

\bibitem[{{Knigge}(2006)}]{Knigge06mn}
{Knigge}, C. 2006, MNRAS, 373, 484

\bibitem[{{Littlefair} {et~al.}(2008){Littlefair}, {Dhillon}, {Marsh},
  {G{\"a}nsicke}, {Southworth}, {Baraffe}, {Watson}, \&
  {Copperwheat}}]{Littlefair+08mn}
{Littlefair}, S.~P., {Dhillon}, V.~S., {Marsh}, T.~R., {et~al.} 2008, MNRAS,
  388, 1582

\bibitem[{{Littlefair} {et~al.}(2006){Littlefair}, {Dhillon}, {Marsh},
  {G{\"a}nsicke}, {Southworth}, \& {Watson}}]{Littlefair+06sci}
{Littlefair}, S.~P., {Dhillon}, V.~S., {Marsh}, T.~R., {et~al.} 2006, Science,
  314, 1578

\bibitem[{{Marsh}(1989)}]{Marsh89pasp}
{Marsh}, T.~R. 1989, PASP, 101, 1032

\bibitem[{{Morrissey} {et~al.}(2007){Morrissey}, {Conrow}, {Barlow}, {Small},
  {Seibert}, {Wyder}, {Budav{\'a}ri}, {Arnouts}, {Friedman}, {Forster}, \&
  {Others}}]{Morrissey+07apjs}
{Morrissey}, P., {Conrow}, T., {Barlow}, T.~A., {et~al.} 2007, ApJS, 173, 682

\bibitem[{{Nair} {et~al.}(2005){Nair}, {Kafka}, {Honeycutt}, \&
  {Gilliland}}]{Nair+05ibvs}
{Nair}, P.~H., {Kafka}, S., {Honeycutt}, R.~K., \& {Gilliland}, R.~L. 2005,
  IBVS, 5585

\bibitem[{{Norton} {et~al.}(2004){Norton}, {Wynn}, \&
  {Somerscales}}]{Norton++04apj}
{Norton}, A.~J., {Wynn}, G.~A., \& {Somerscales}, R.~V. 2004, ApJ, 614, 349

\bibitem[{{Patterson}(1998)}]{Patterson98pasp}
{Patterson}, J. 1998, PASP, 110, 1132

\bibitem[{{Ritter} \& {Kolb}(2003)}]{RitterKolb03aa}
{Ritter}, H. \& {Kolb}, U. 2003, A\&A, 404, 301

\bibitem[{{Rodr{\'{\i}}guez-Gil} {et~al.}(2007){Rodr{\'{\i}}guez-Gil},
  {G{\"a}nsicke}, {Hagen}, {Araujo-Betancor}, {Aungwerojwit}, {Allende Prieto},
  {Boyd}, {Casares}, {Engels}, \& {Giannakis}}]{Rodriguez+07mn2}
{Rodr{\'{\i}}guez-Gil}, P., {G{\"a}nsicke}, B.~T., {Hagen}, H.-J., {et~al.}
  2007, MNRAS, 377, 1747

\bibitem[{{Scargle}(1982)}]{Scargle82apj}
{Scargle}, J.~D. 1982, ApJ, 263, 835

\bibitem[{{Schmidt} {et~al.}(2008){Schmidt}, {Smith}, {Szkody}, \&
  {Anderson}}]{Schmidt+08pasp}
{Schmidt}, G.~D., {Smith}, P.~S., {Szkody}, P., \& {Anderson}, S.~F. 2008,
  PASP, 120, 160

\bibitem[{{Schmidt} {et~al.}(2005){Schmidt}, {Szkody}, {Vanlandingham},
  {Anderson}, {Barentine}, {Brewington}, {Hall}, {Harvanek}, {Kleinman}, \&
  {Others}}]{Schmidt+05apj}
{Schmidt}, G.~D., {Szkody}, P., {Vanlandingham}, K.~M., {et~al.} 2005, ApJ,
  630, 1037

\bibitem[{{Schneider} \& {Young}(1980)}]{SchneiderYoung80apj}
{Schneider}, D.~P. \& {Young}, P. 1980, ApJ, 238, 946

\bibitem[{{Schwarzenberg-Czerny}(1989)}]{Schwarzenberg89mn}
{Schwarzenberg-Czerny}, A. 1989, MNRAS, 241, 153

\bibitem[{{Schwarzenberg-Czerny}(1996)}]{Schwarzenberg96apj}
{Schwarzenberg-Czerny}, A. 1996, ApJ, 460, L107

\bibitem[{{Shears} {et~al.}(2011){Shears}, {Brady}, {Campbell}, {Henden}, {de
  Miguel}, {Morelle}, {Roberts}, {Sabo}, \& {Miller}}]{Shears+11xxx}
{Shears}, J., {Brady}, S., {Campbell}, T., {et~al.} 2011, JBAA accepted, {\it
  arXiv:1104.0104}

\bibitem[{{Southworth} {et~al.}(2010){Southworth}, {Copperwheat}, {Gansicke},
  \& {Pyrzas}}]{Me+10aa}
{Southworth}, J., {Copperwheat}, C.~M., {Gansicke}, B., \& {Pyrzas}, S. 2010,
  A\&A, 510, A100

\bibitem[{{Southworth} {et~al.}(2012){Southworth}, {G{\"a}nsicke}, \&
  {Breedt}}]{Me++12iaus}
{Southworth}, J., {G{\"a}nsicke}, B.~T., \& {Breedt}, E. 2012, in IAU Symposium
  282, Cambridge University Press, Cambridge, UK., ed. M.~T. {Richards} \&
  I.~{Hubeny}, 123--124

\bibitem[{{Southworth} {et~al.}(2007{\natexlab{a}}){Southworth},
  {G{\"a}nsicke}, {Marsh}, {de Martino}, \& {Aungwerojwit}}]{Me+07mn}
{Southworth}, J., {G{\"a}nsicke}, B.~T., {Marsh}, T.~R., {de Martino}, D., \&
  {Aungwerojwit}, A. 2007{\natexlab{a}}, MNRAS, 378, 635

\bibitem[{{Southworth} {et~al.}(2006){Southworth}, {G{\"a}nsicke}, {Marsh}, {de
  Martino}, {Hakala}, {Littlefair}, {Rodr{\'{\i}}guez-Gil}, \&
  {Szkody}}]{Me+06mn}
{Southworth}, J., {G{\"a}nsicke}, B.~T., {Marsh}, T.~R., {et~al.} 2006, MNRAS,
  373, 687

\bibitem[{{Southworth} {et~al.}(2008{\natexlab{a}}){Southworth},
  {G{\"a}nsicke}, {Marsh}, {Torres}, {Steeghs}, {Hakala}, {Copperwheat},
  {Aungwerojwit}, \& {Mukadam}}]{Me+08mn}
{Southworth}, J., {G{\"a}nsicke}, B.~T., {Marsh}, T.~R., {et~al.}
  2008{\natexlab{a}}, MNRAS, 391, 591

\bibitem[{{Southworth} {et~al.}(2009{\natexlab{a}}){Southworth}, {Hickman},
  {Marsh}, {Rebassa-Mansergas}, {G{\"a}nsicke}, {Copperwheat}, \&
  {Rodr{\'{\i}}guez-Gil}}]{Me+09aa}
{Southworth}, J., {Hickman}, R.~D.~G., {Marsh}, T.~R., {et~al.}
  2009{\natexlab{a}}, A\&A, 507, 929

\bibitem[{{Southworth} {et~al.}(2014){Southworth}, {Hinse}, {Burgdorf}, {Calchi
  Novati}, {Dominik}, {Galianni}, {Gerner}, {Giannini}, {Gu}, {Hundertmark},
  {J{\o}rgensen}, {Juncher}, {Kerins}, {Mancini}, {Rabus}, {Ricci},
  {Sch{\"a}fer}, {Skottfelt}, {Tregloan-Reed}, {Wang}, {Wertz}, {Alsubai},
  {Andersen}, {Bozza}, {Bramich}, {Browne}, {Ciceri}, {D'Ago}, {Damerdji},
  {Diehl}, {Dodds}, {Elyiv}, {Fang}, {Finet}, {Figuera Jaimes}, {Hardis},
  {Harps{\o}e}, {Jessen-Hansen}, {Kains}, {Kjeldsen}, {Korhonen}, {Liebig},
  {Lund}, {Lundkvist}, {Mathiasen}, {Penny}, {Popovas}, {Prof.}, {Rahvar},
  {Sahu}, {Scarpetta}, {Schmidt}, {Sch{\"o}nebeck}, {Snodgrass}, {Street},
  {Surdej}, {Tsapras}, \& {Vilela}}]{Me+14mn}
{Southworth}, J., {Hinse}, T.~C., {Burgdorf}, M., {et~al.} 2014, MNRAS, 444,
  776

\bibitem[{{Southworth} {et~al.}(2009{\natexlab{b}}){Southworth}, {Hinse},
  {J{\o}rgensen}, {Dominik}, {Ricci}, {Burgdorf}, {Hornstrup}, {Wheatley},
  {Anguita}, {Bozza}, {Calchi Novati}, {Harps{\o}e}, {Kj{\ae}rgaard}, {Liebig},
  {Mancini}, {Masi}, {Mathiasen}, {Rahvar}, {Scarpetta}, {Snodgrass}, {Surdej},
  {Th{\"o}ne}, \& {Zub}}]{Me+09mn}
{Southworth}, J., {Hinse}, T.~C., {J{\o}rgensen}, U.~G., {et~al.}
  2009{\natexlab{b}}, MNRAS, 396, 1023

\bibitem[{{Southworth} {et~al.}(2007{\natexlab{b}}){Southworth}, {Marsh},
  {G{\"a}nsicke}, {Aungwerojwit}, {Hakala}, {de Martino}, \&
  {Lehto}}]{Me+07mn2}
{Southworth}, J., {Marsh}, T.~R., {G{\"a}nsicke}, B.~T., {et~al.}
  2007{\natexlab{b}}, MNRAS, 382, 1145

\bibitem[{{Southworth} {et~al.}(2008{\natexlab{b}}){Southworth}, {Townsley}, \&
  {G{\"a}nsicke}}]{Me++08mn}
{Southworth}, J., {Townsley}, D.~M., \& {G{\"a}nsicke}, B.~T.
  2008{\natexlab{b}}, MNRAS, 388, 709

\bibitem[{{Stetson}(1987)}]{Stetson87pasp}
{Stetson}, P.~B. 1987, PASP, 99, 191

\bibitem[{{Szkody} {et~al.}(2011){Szkody}, {Anderson}, {Brooks},
  {G{\"a}nsicke}, {Kronberg}, {Riecken}, {Ross}, {Schmidt}, {Schneider},
  {Ag{\"u}eros}, {Gomez-Moran}, {Knapp}, {Schreiber}, \&
  {Schwope}}]{Szkody+11aj}
{Szkody}, P., {Anderson}, S.~F., {Brooks}, K., {et~al.} 2011, AJ, 142, 181

\bibitem[{{Szkody} {et~al.}(2009){Szkody}, {Anderson}, {Hayden}, {Kronberg},
  {McGurk}, {Riecken}, {Schmidt}, {West}, {G{\"a}nsicke}, {Nebot Gomez-Moran},
  {Schneider}, {Schreiber}, \& {Schwope}}]{Szkody+09aj}
{Szkody}, P., {Anderson}, S.~F., {Hayden}, M., {et~al.} 2009, AJ, 137, 4011

\bibitem[{{Szkody} {et~al.}(2003){Szkody}, {Anderson}, {Schmidt}, {Hall},
  {Margon}, {Miceli}, {SubbaRao}, {Frith}, {Harris}, {Hawley}, \&
  {Others}}]{Szkody+03apj}
{Szkody}, P., {Anderson}, S.~F., {Schmidt}, G., {et~al.} 2003, ApJ, 583, 902

\bibitem[{{Szkody} \& {Brownlee}(1977)}]{SzkodyBrownlee77apj}
{Szkody}, P. \& {Brownlee}, D.~E. 1977, ApJ, 212, L113

\bibitem[{{Szkody} {et~al.}(2004){Szkody}, {Henden}, {Fraser}, {Silvestri},
  {Bochanski}, {Wolfe}, {Ag{\"u}eros}, {Warner}, {Woudt}, {Tramposch}, \&
  {Others}}]{Szkody+04aj}
{Szkody}, P., {Henden}, A., {Fraser}, O., {et~al.} 2004, AJ, 128, 1882

\bibitem[{{Szkody} {et~al.}(2005){Szkody}, {Henden}, {Fraser}, {Silvestri},
  {Schmidt}, {Bochanski}, {Wolfe}, {Ag{\"u}eros}, {Anderson}, {Mannikko},
  {Downes}, {Schneider}, \& {Brinkmann}}]{Szkody+05aj}
{Szkody}, P., {Henden}, A., {Fraser}, O.~J., {et~al.} 2005, AJ, 129, 2386

\bibitem[{{Szkody} {et~al.}(2007){Szkody}, {Henden}, {Mannikko}, {Mukadam},
  {Schmidt}, {Bochanski}, {Ag{\"u}eros}, {Anderson}, {Silvestri}, {Dahab},
  {Oguri}, {Schneider}, {Shin}, {Strauss}, {Knapp}, \& {West}}]{Szkody+07aj}
{Szkody}, P., {Henden}, A., {Mannikko}, L., {et~al.} 2007, AJ, 134, 185

\bibitem[{{Thorstensen} {et~al.}(1991){Thorstensen}, {Ringwald}, {Wade},
  {Schmidt}, \& {Norsworthy}}]{Thorstensen+91aj}
{Thorstensen}, J.~R., {Ringwald}, F.~A., {Wade}, R.~A., {Schmidt}, G.~D., \&
  {Norsworthy}, J.~E. 1991, AJ, 102, 272

\bibitem[{{Thorstensen} \& {Skinner}(2012)}]{ThorstensenSkinner12aj}
{Thorstensen}, J.~R. \& {Skinner}, J.~N. 2012, AJ, 144, 81

\bibitem[{{Tovmassian} {et~al.}(2014){Tovmassian}, {Stephania Hernandez},
  {Gonz{\'a}lez-Buitrago}, {Zharikov}, \&
  {Garc{\'{\i}}a-D{\'{\i}}az}}]{Tovmassian+14aj}
{Tovmassian}, G., {Stephania Hernandez}, M., {Gonz{\'a}lez-Buitrago}, D.,
  {Zharikov}, S., \& {Garc{\'{\i}}a-D{\'{\i}}az}, M.~T. 2014, AJ, 147, 68

\bibitem[{{Uemura} {et~al.}(2010){Uemura}, {Kato}, {Nogami}, \&
  {Ohsugi}}]{Uemura10pasj}
{Uemura}, M., {Kato}, T., {Nogami}, D., \& {Ohsugi}, T. 2010, PASJ, 62, 613

\bibitem[{{Warner}(1995)}]{Warner95book}
{Warner}, B. 1995, {Cataclysmic Variable Stars} (Cambridge Astrophysics Series,
  Cambridge University Press, Cambridge, UK)

\bibitem[{{Warner} \& {Nather}(1972)}]{WarnerNather72}
{Warner}, B. \& {Nather}, R.~E. 1972, MNRAS, 156, 305

\bibitem[{{Webbink} \& {Wickramasinghe}(2002)}]{WebbinkWick02mn}
{Webbink}, R.~F. \& {Wickramasinghe}, D.~T. 2002, MNRAS, 335, 1

\bibitem[{{Whyte} \& {Eggleton}(1980)}]{WhyteEggleton80mn}
{Whyte}, C.~A. \& {Eggleton}, P.~P. 1980, MNRAS, 190, 801

\bibitem[{{Wils} {et~al.}(2010){Wils}, {G{\"a}nsicke}, {Drake}, \&
  {Southworth}}]{Wils+10mn}
{Wils}, P., {G{\"a}nsicke}, B.~T., {Drake}, A.~J., \& {Southworth}, J. 2010,
  MNRAS, 402, 436

\bibitem[{{Wood} {et~al.}(1989){Wood}, {Horne}, {Berriman}, \&
  {Wade}}]{Wood+89apj}
{Wood}, J.~H., {Horne}, K., {Berriman}, G., \& {Wade}, R.~A. 1989, ApJ, 341,
  974

\bibitem[{{Woudt} {et~al.}(2012){Woudt}, {Warner}, {de Bud{\'e}}, {Macfarlane},
  {Schurch}, \& {Zietsman}}]{Woudt+12mn}
{Woudt}, P.~A., {Warner}, B., {de Bud{\'e}}, D., {et~al.} 2012, MNRAS, 421,
  2414

\end{thebibliography}

\end{document}